\documentclass[american,english]{svjour3}
\usepackage[T1]{fontenc}
\usepackage[latin9]{inputenc}
\usepackage{geometry}
\usepackage{color}
\usepackage{array}
\usepackage{longtable}
\usepackage{float}
\usepackage{url}
\usepackage{algorithm2e}
\usepackage{amsmath}
\usepackage{amssymb}
\usepackage{graphicx}
\usepackage{setspace}
\usepackage[authoryear]{natbib}

\makeatletter

\providecommand{\tabularnewline}{\\}

  \newenvironment{svmultproof}{\begin{proof}}{\qed\end{proof}}

\usepackage{mcode}

\usepackage{babel}
\usepackage{listings}
\usepackage[labelfont=bf,labelsep=space]{caption}

\renewcommand{\fnum@figure}{\textbf{Fig.~\thefigure}}

\addto\captionsamerican{}
\addto\captionsenglish{}

\makeatother

  \addto\captionsamerican{}
  \addto\captionsamerican{}
  \addto\captionsamerican{}
  \addto\captionsenglish{}
  \addto\captionsenglish{}
  \addto\captionsenglish{}

\addto\captionsamerican{}
\addto\captionsenglish{}

\begin{document}

\title{Random Neighborhood Graphs as Models of Fracture Networks on Rocks:
Structural and Dynamical Analysis}

\author{Ernesto Estrada and Matthew Sheerin}

\institute{Department of Mathematics \& Statistics, University of Strathclyde,
26 Richmond Street, Glasgow G1 1XQ}
\maketitle
\begin{abstract}
We propose a new model to account for the main structural characteristics
of rock fracture networks (RFNs). The model is based on a generalization
of the random neighborhood graphs to consider fractures embedded into
rectangular spaces. We study a series of 29 real-world RFNs and find
the best fit with the random rectangular neighborhood graphs (RRNGs)
proposed here. We show that this model captures most of the structural
characteristics of the RFNs and allows a distinction between small
and more spherical rocks and large and more elongated ones. We use
a diffusion equation on the graphs in order to model diffusive processes
taking place through the channels of the RFNs. We find a small set
of structural parameters that highly correlates with the average diffusion
time in the RFNs. We found analytically some bounds for the diameter
and the algebraic connectivity of these graphs that allow to bound
the diffusion time in these networks. We also show that the RRNGs
can be used as a suitable model to replace the RFNs in the study of
diffusion-like processes. Indeed, the diffusion time in RFNs can be
predicted by using structural and dynamical parameters of the RRNGs.
Finally, we also explore some potential extensions of our model to
include variable fracture apertures, the possibility of long-range
hops of the diffusive particles as a way to account for heterogeneities
in the medium and possible superdiffusive processes, and the extension
of the model to 3-dimensional space.
\end{abstract}

\section{Introduction}

It could be argued that in any system transporting mass and energy
there should be an underlying network responsible for conducting the
materials through space. The flow of fluids of petrochemical interest
obeys this general rule, where the role of the transporting network
is played by the system of rock fractures. The study of rock fracture
systems have a long tradition in hydrocarbon geology and hydrogeology
due to the role that these fractures play on the evaluation of potential
oil reservoirs (\cite{Fractures_1,Fractures_2,Fractures_3,Fractures_4,Fractures_5,Fractures_6}).
The analysis of rock fracture networks (RFNs) plays a fundamental
role in determining the nature and disposition of heterogeneities
appearing in petroliferous formations to determine the capability
for the transport of fluid through them (\cite{Reservoir_1,Reservoir_2,Reservoir_3,Reservoir_4,Reservoir_5}).
In many of the analyses described in the literature the use of synthetic
fracture networks facilitates the analysis due to the sometimes scarce
availability of real-world data (\cite{Synthetic_1,Synthetic_2,Synthetic_3,Synthetic_4,Synthetic_5,Synthetic_6,Synthetic_7}).
In contrast, Santiago et al. have published a series of papers (\cite{Santiago_1,Santiago_centralities,Santiago_metrics})
in which they used real fracture networks derived from original hand-sampled
images of rocks extracted from a Gulf of Mexico oil reservoir. These
works have used a graph-theoretic analysis of these real-world networks
in order to extract information about the topological (static) characteristics
of this group of rocks. Rock fractures have also been studied in a
more general sense for their applications to both oil reservoirs and
other fluid flows within rocks such as groundwater, examining properties
such as fractal scaling and anomalous diffusion (\cite{berkowitz_2,berkowitz_1,berkowitz_3}).

This work is a step forward in the analysis of real rock fracture
networks. First, a synthetic model reproducing the topology of real
RFNs is proposed. This model, which is based on a generalization of
the random neighborhood graphs (\cite{Beta_skeleton,Beta_skeletons_jaromczyk}),
is compared statistically with the real RFNs using a thorough topological
characterization of the structure of these networks. The random neighborhood
graphs represent a family of simple graphs in which two vertices are
connected by an edge if and only if the vertices satisfy particular
geometric requirements, and they involve a spatial distance. The Euclidean
distance is most common choice and will be used here. They have found
multiple applications in cases where spatial properties of graphs
are required.

A discrete version of the diffusion equation is used to study the
diffusion of a fluid through the channels of the real RFNs. The diffusion
through the real RFNs is compared to diffusion on the synthetic model,
showing that this random model reproduces not only the most important
structural properties of the real networks but also their diffusive
properties. As in the series of papers by \cite{Santiago_1,Santiago_centralities,Santiago_metrics},
two-dimensional cuts of rocks that show a fracture network embedded
into the rock sides are considered. Then, a potential criticism to
these works is the fact that rock networks are three-dimensional (\cite{3D_fractures_1,3D_fractures_2,3D_fractures_3})
and that inferring these 3D networks from 2D information is hard.
However, as has been previously documented, the analysis of 2D rock
fractures identifies important parameters that allow the characterization
of real rock samples (\cite{2D_analysis_1,2D_analysis_2,Santiago_1,Santiago_centralities,Santiago_metrics}).
In addition, note that the generalized proximity graphs that are introduced
in this work can be easily extended to the 3D case. Thus, 3D rock
fracture networks extrapolating the topological information that is
obtained here from the analysis of 2D samples can be easily generated.

\section{Rectangular $\beta$-Skeleton Graphs and Relative Neighborhood Graphs}

\begin{figure}[H]
\begin{centering}
\begin{tabular}{|>{\centering}m{1.5cm}|>{\centering}m{6cm}|>{\centering}m{7cm}|}
\hline 
$\beta$ Value  & Construction of $\beta$-skeleton  & Example of $\beta$-skeleton\tabularnewline
\hline 
$\beta<1$  & \includegraphics[width=0.19\paperheight]{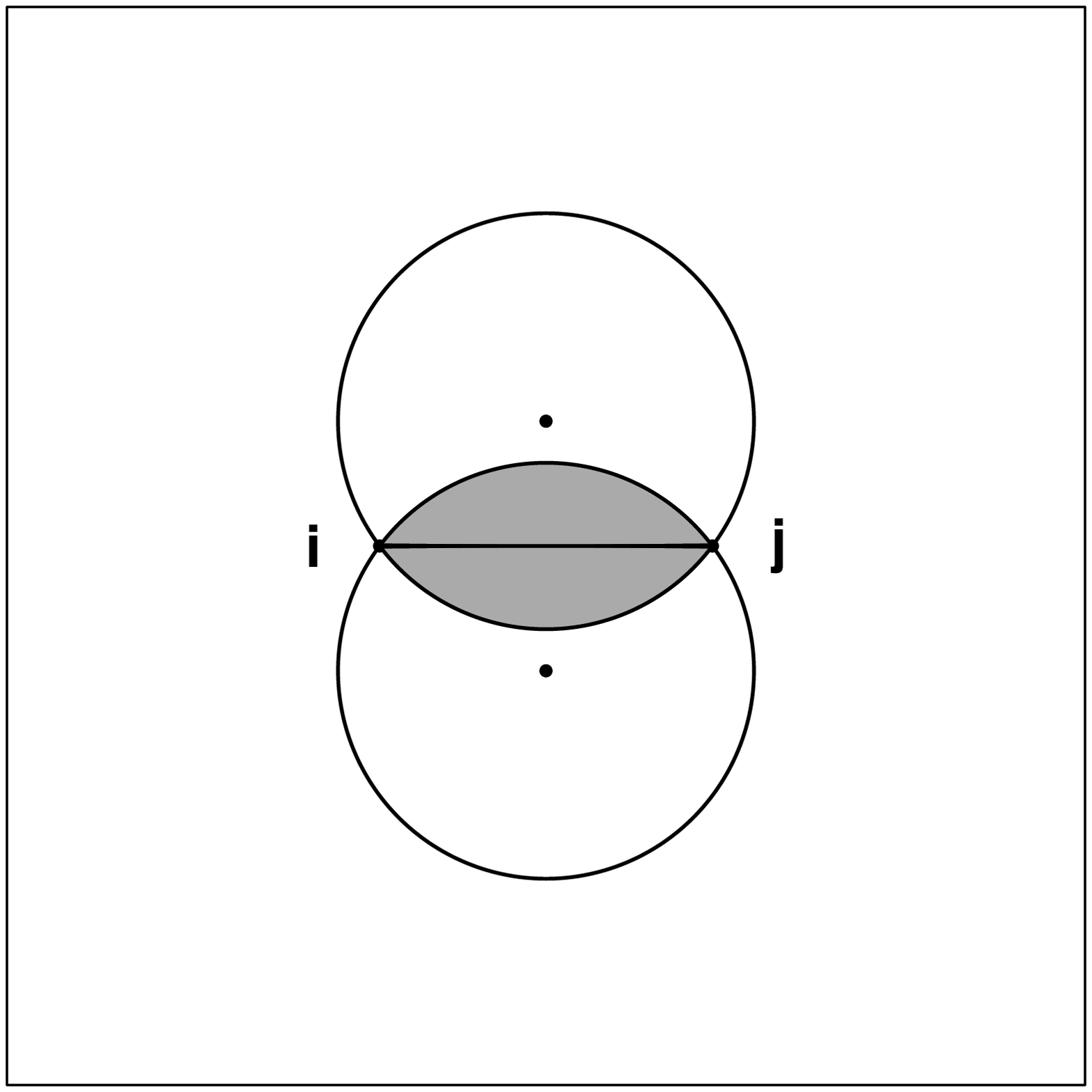}  & \includegraphics[width=0.19\paperheight]{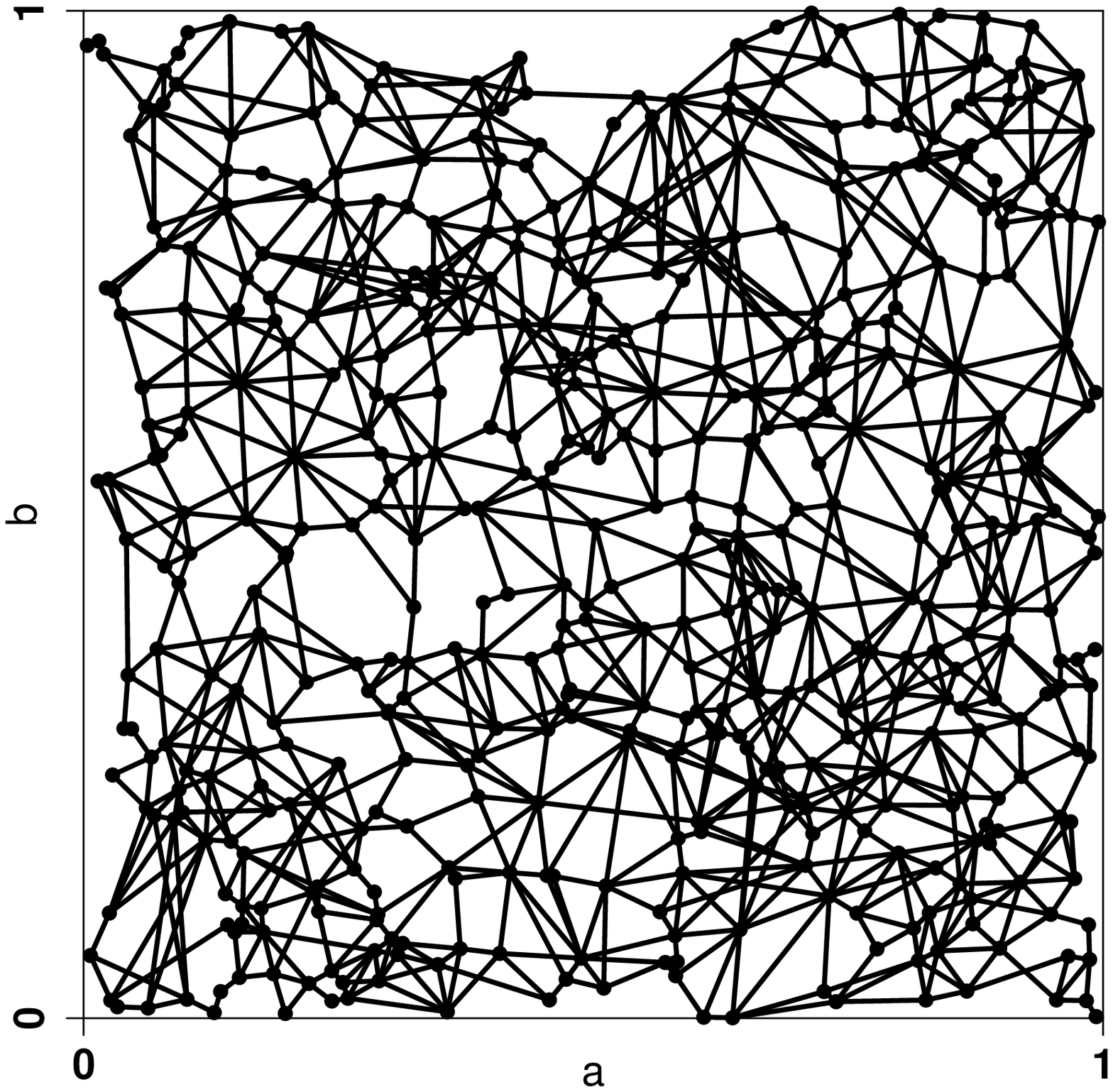}\tabularnewline
\hline 
$\beta=1$  & \includegraphics[width=0.19\paperheight]{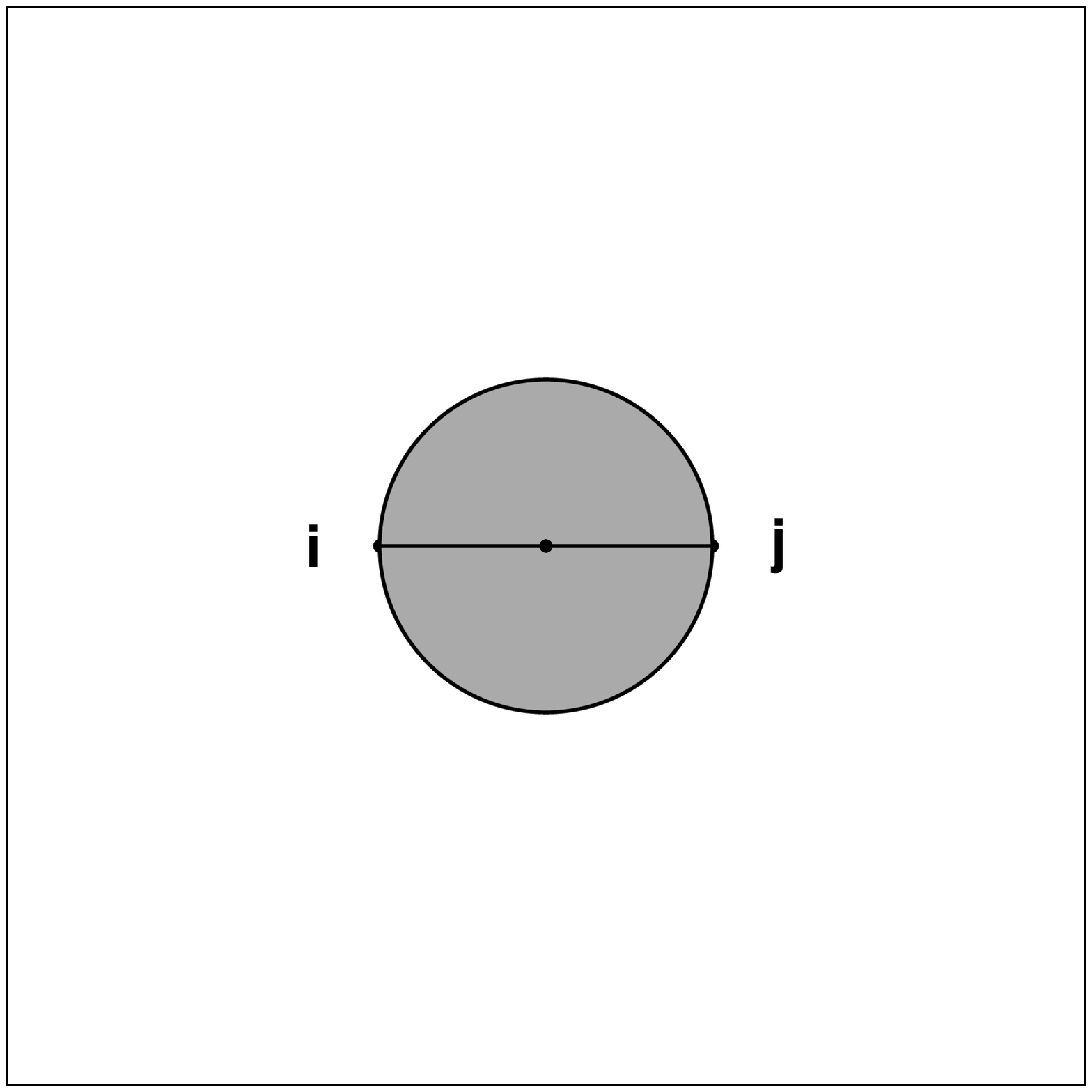}  & \includegraphics[width=0.19\paperheight]{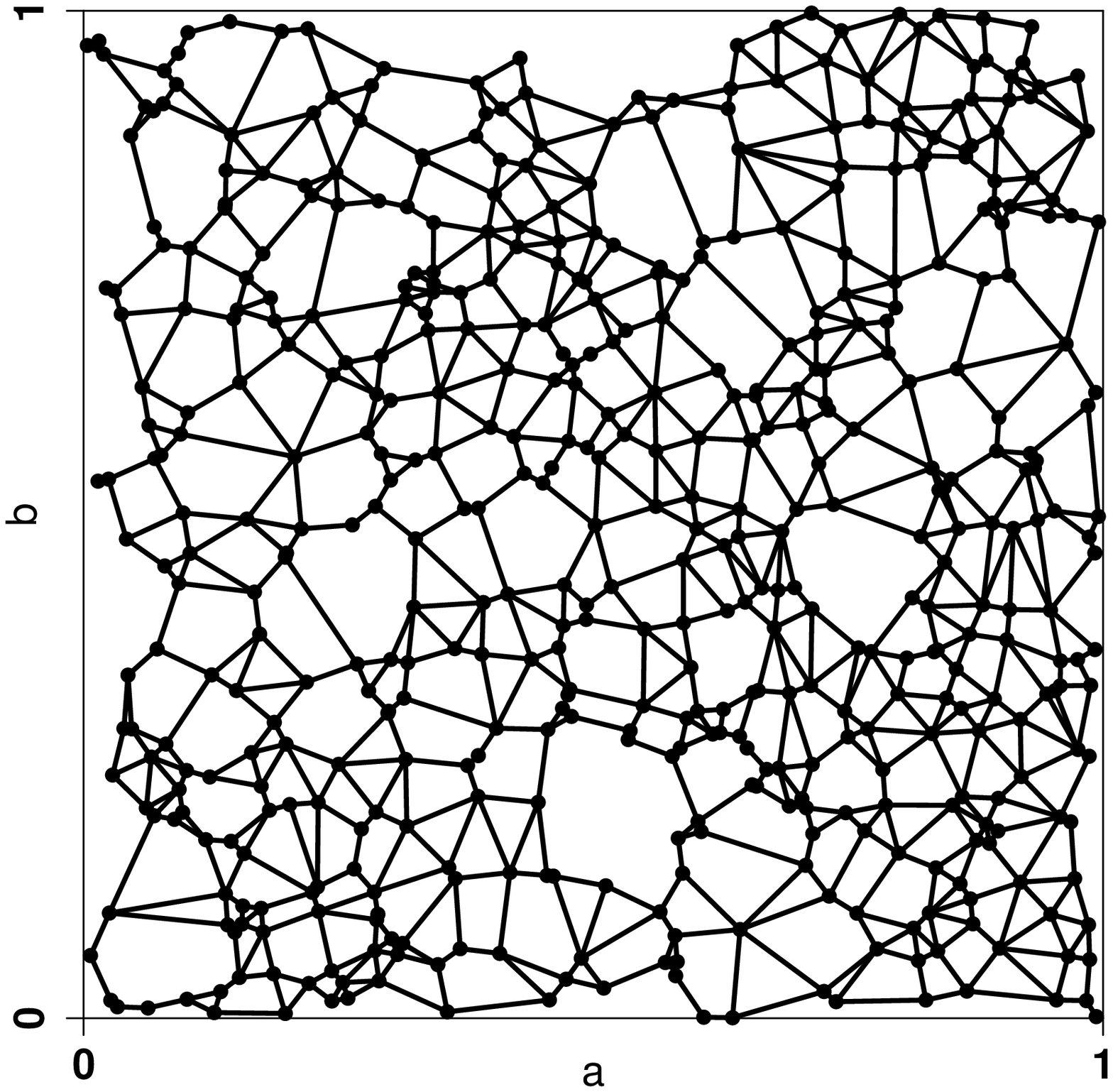}\tabularnewline
\hline 
$1<\beta<2$  & \includegraphics[width=0.19\paperheight]{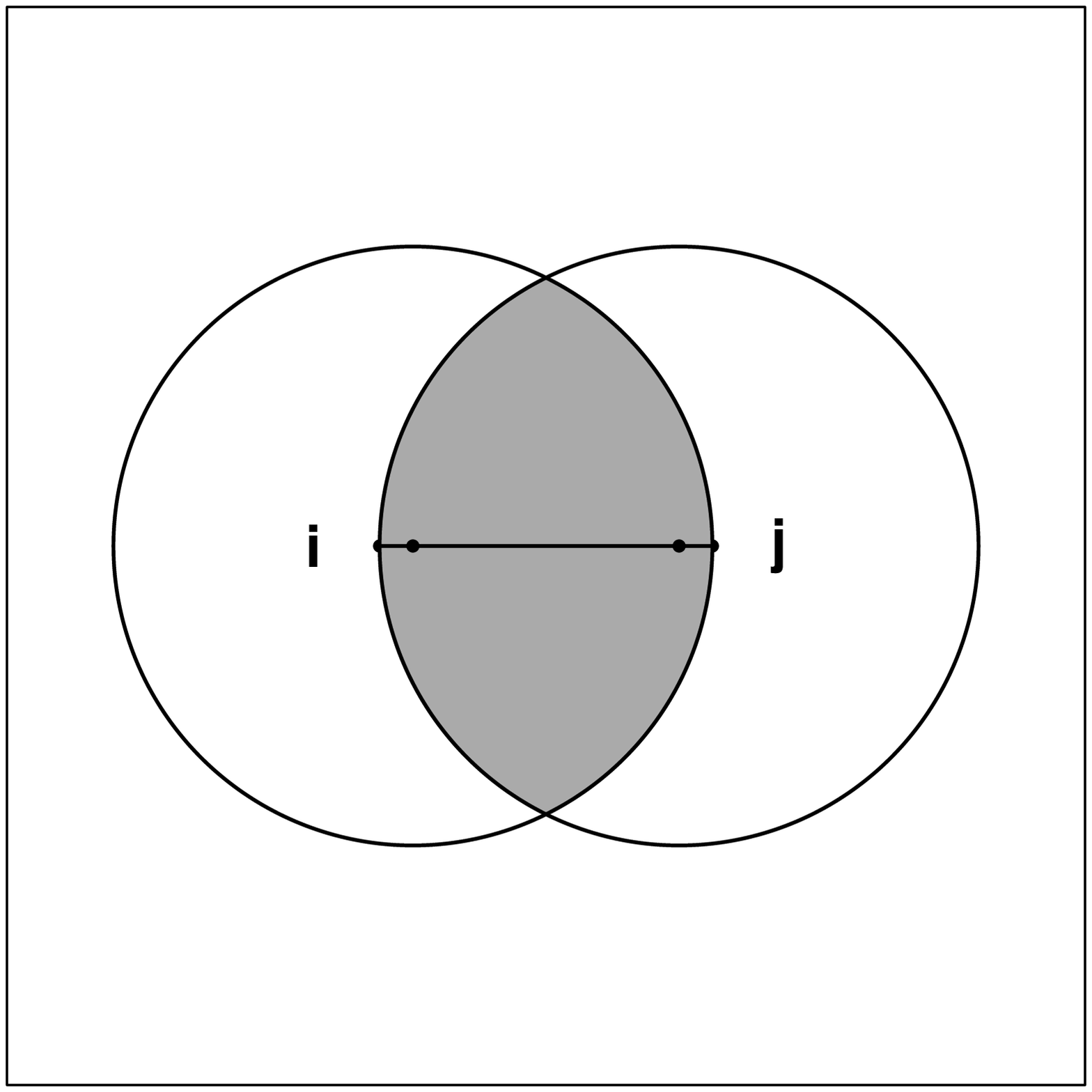}  & \includegraphics[width=0.19\paperheight]{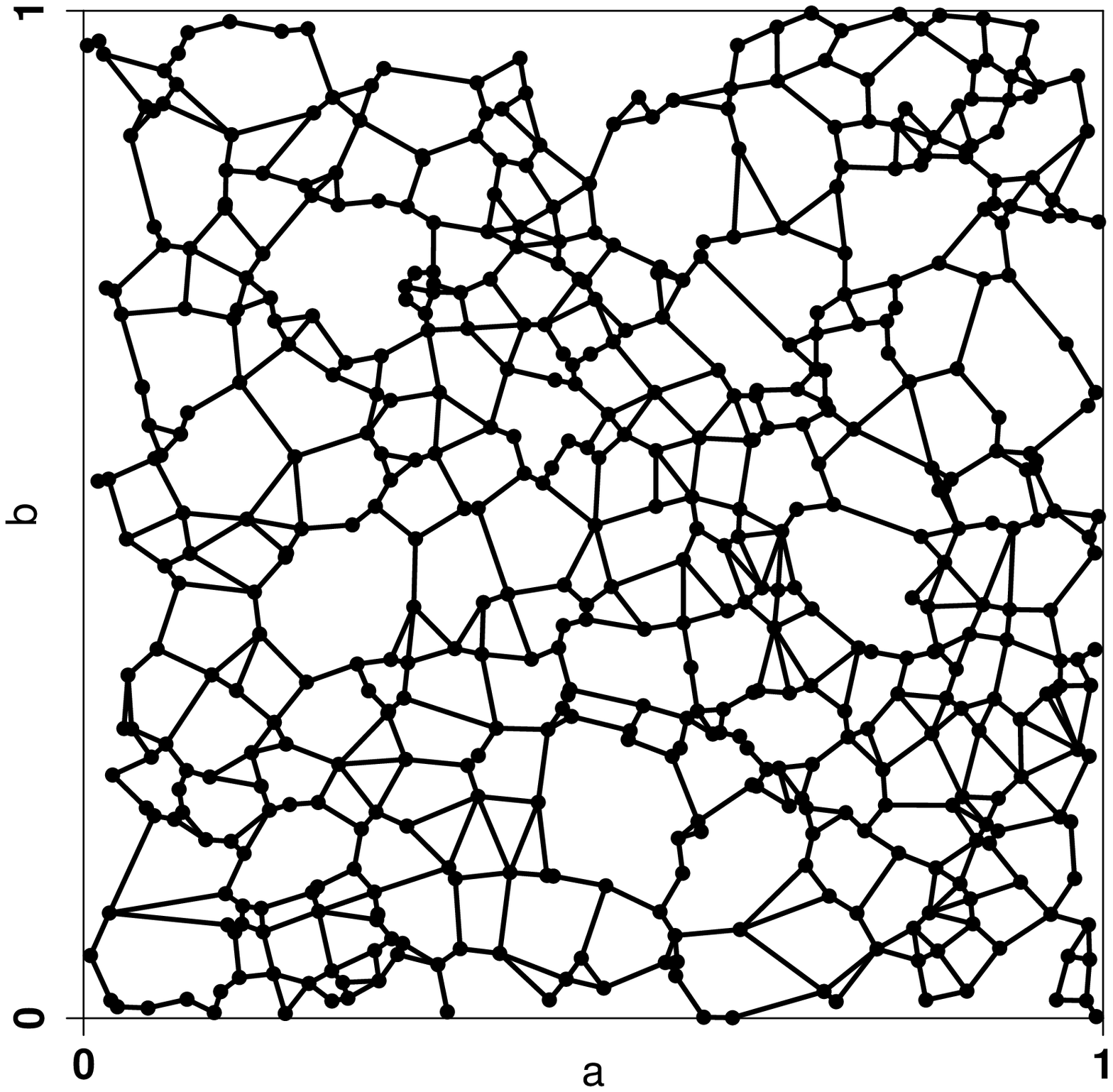}\tabularnewline
\hline 
$\beta=2$  & \includegraphics[width=0.19\paperheight]{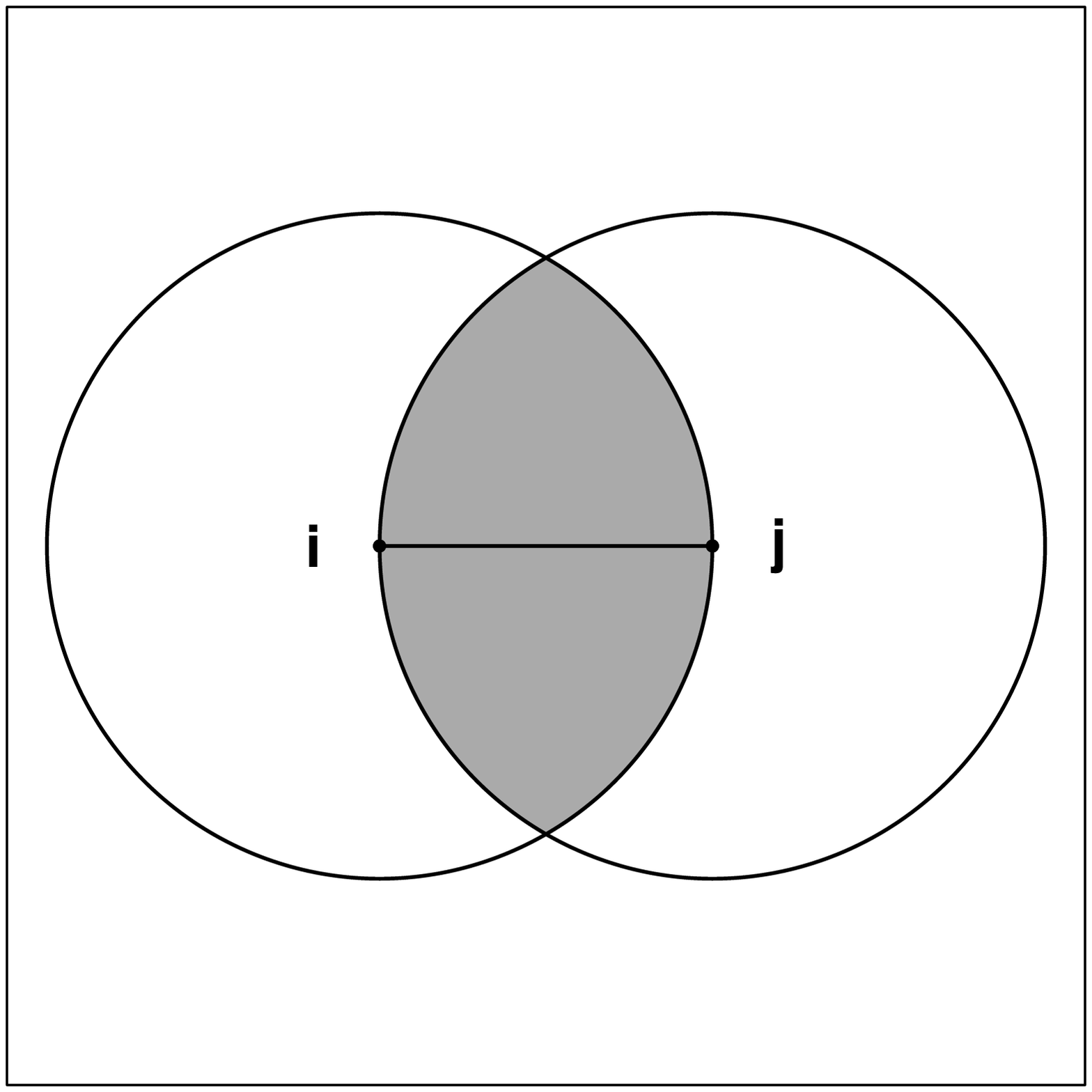}  & \includegraphics[width=0.19\paperheight]{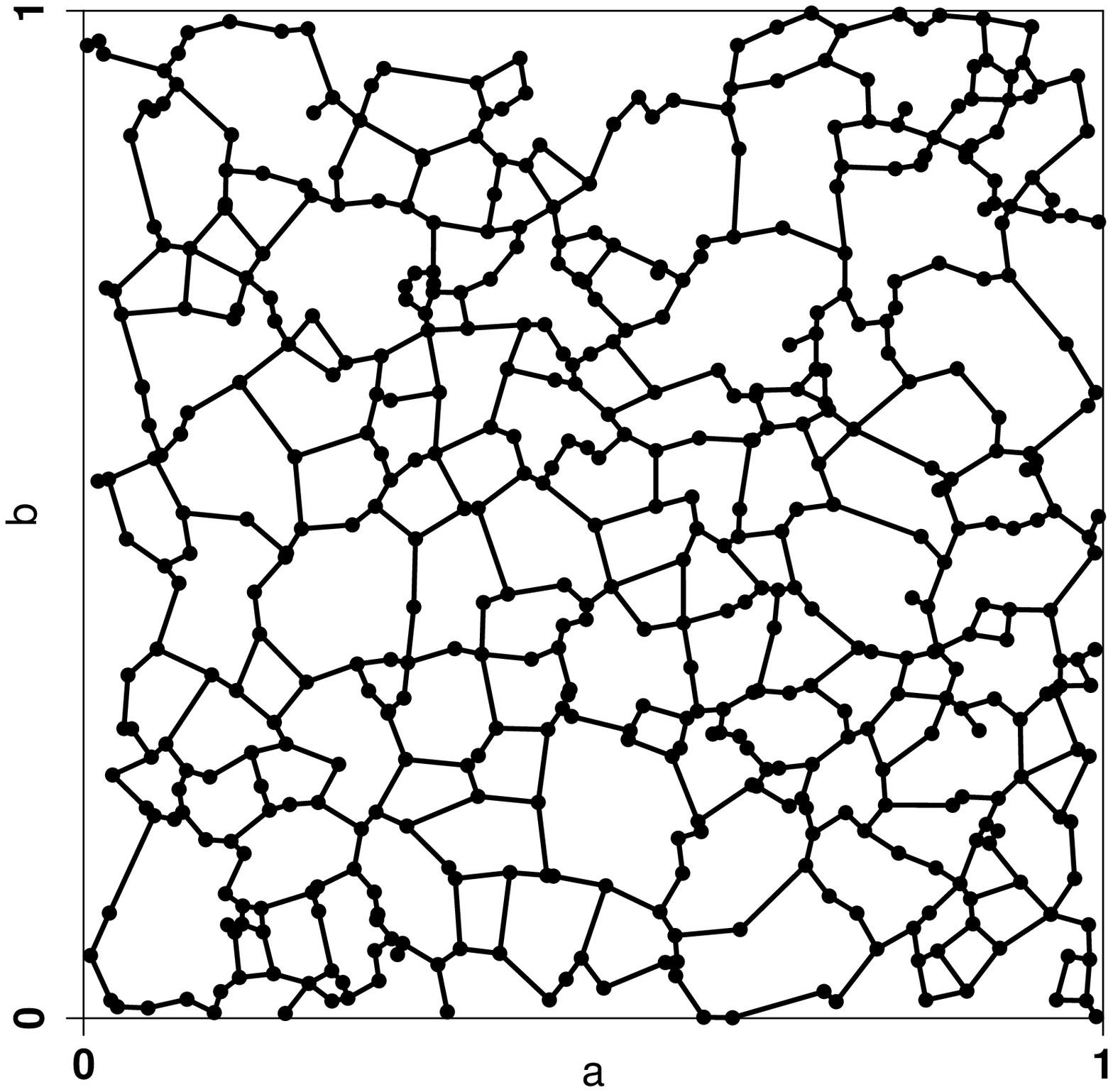}\tabularnewline
\hline 
\end{tabular}
\par\end{centering}
\caption{Illustration of the construction of $\beta$-skeletons in a unit square
for different values of $\beta$, where the region $R$ is shaded
gray, with an example with $n=500$ nodes for each. From top to bottom:
$\beta=0.8,1.0,1.8,2.0$. }
\label{beta_skeletons_examples} 
\end{figure}

This section describes a generalization of the so-called $\beta$-skeleton
graphs by considering not only points randomly distributed in a unit
square but also in unit rectangles. The `classical' $\beta$-skeleton
graphs are described by \cite{Beta_skeletons_jaromczyk,Beta_skeleton,rock_fractures}.
The model can be briefly described as follows. Consider $n$ points
$p_{i}$ $\left(i=1,2,\ldots,n\right)$ randomly and independently
distributed in a unit square, and a value $\beta\geq1$. Let $p_{i}$
and $p_{j}$ be two arbitrary points which are separated by a Euclidean
distance $L$, and let $B(x,r)$ denote the circle located at $x$
with radius $r$. For $\beta\geq1$ the lune-based definition of the
$\beta$-skeleton model is used and described here as it is more suitable
for our needs, an alternative is the circle-based definition which
we do not examine here. Two circles, $B((1-\frac{\beta}{2})p_{i}+\frac{\beta}{2}p_{j},\frac{\beta}{2}L)$
and $B((1-\frac{\beta}{2})p_{j}+\frac{\beta}{2}p_{i},\frac{\beta}{2}L)$
are constructed, and let $R$ be intersection of the circles. It is
obvious that the area of $R$ increases as $\beta$ is increased.
Although it was previously stated that $\beta\geq1$ for the sake
of the current paper, the case of $0<\beta<1$ is also defined. In
this case two circles of radius $L/(2\beta)$ which pass through both
points $p_{i}$ and $p_{j}$ are instead constructed, and again the
intersection is denoted by $R$, and note that the construction of
the circles differs from the case of $\beta\geq1$. Then, if there
is no other point $p_{k}$ included in the region $R$, the points
$p_{i}$ and $p_{j}$ are connected by a segment of line, otherwise
the points are not connected. By considering this process for all
pairs of points, a graph $G=\left(V,E\right)$ is constructed in which
the set of vertices $V$ is formed by the points $p_{i}$ and the
set of edges $E$ is formed by the segments of lines connecting pairs
of vertices. Obviously, for small values of $\beta$, e.g., for $0<\beta<1$,
the chances that there is a point in the region $R$ associated with
$p_{i}$ and $p_{j}$ is very small, and there is a high probability
that these two points are connected (see Fig.~\ref{beta_skeletons_examples}).
As a consequence of this, the resulting graphs are very dense, containing
a large number of triangles. It can be seen that if $\beta=0$, the
resulting graph is just the complete graph. Another particular case
which is commonly considered in the computational geometry literature
is when $\beta=2$, which corresponds to the so-called relative neighborhood
graph (RNG). Matlab code for creating such proximity graphs is provided
in Appendix~\ref{appendix:beta_skeletons}.

\subsection{Generalization of $\beta$-skeleton graphs}

The generalization of the $\beta$-skeleton graphs is constructed
by considering a rectangle $[0,a]\times[0,b]$ where $a,b\in\mathbb{R},\ a\geq b$.
Only unit rectangles of the form $[0,a]\times[0,a^{-1}]$ will be
considered here. The rest of the construction of a rectangular $\beta$-skeleton
graph is similar to that of a $\beta$-skeleton graph. That is, $n$
points are distributed uniformly and independently in the unit rectangle
$[0,a]\times[0,a^{-1}]$. Obviously, when $a=1$ the rectangle $[0,a]\times[0,a^{-1}]$
is simply the unit square $[0,1]^{2}$, which means that the rectangular
$\beta$-skeleton graph becomes the classical $\beta$-skeleton one.

\begin{figure}[H]
\begin{centering}
\includegraphics[width=0.5\textwidth]{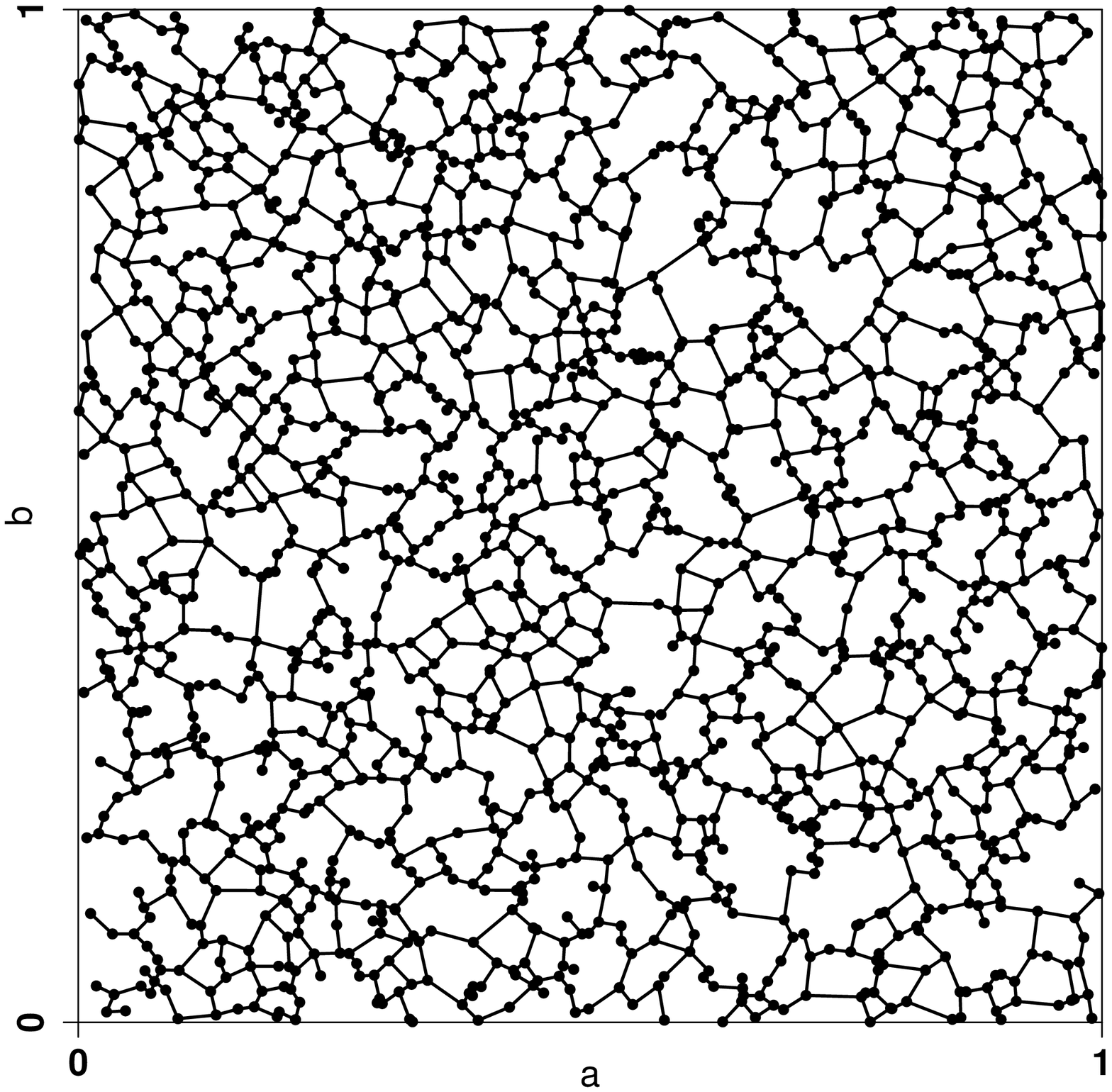} \includegraphics[width=1\textwidth]{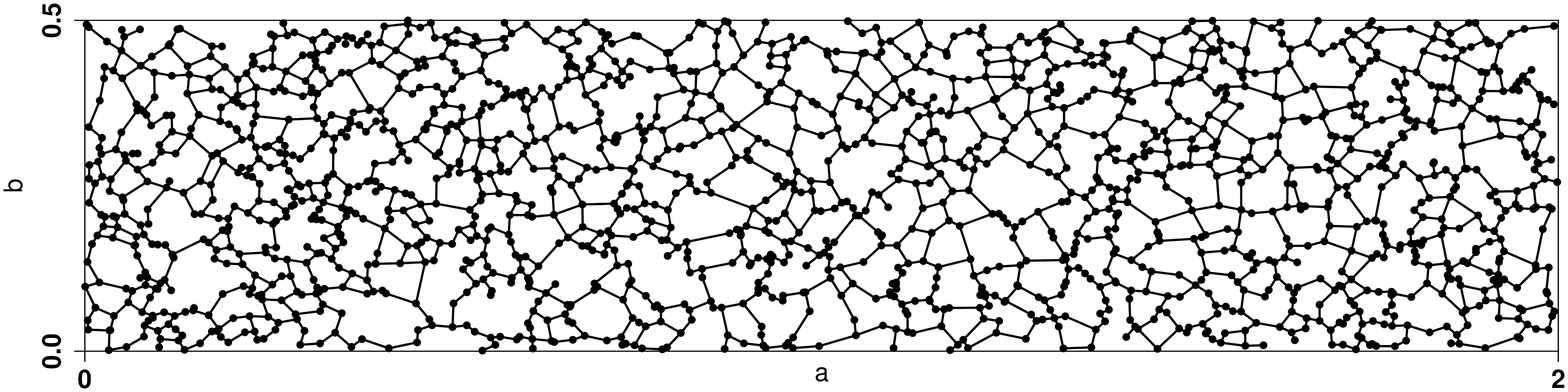} 
\par\end{centering}
\caption{Example RNGs for $a=1$ (top) and $a=2$ (bottom), with $n=1686$
nodes which corresponds to a particular RFN.}
\label{rectangular beta skeleton} 
\end{figure}

Figure \ref{rectangular beta skeleton} illustrates two rectangular
$\beta$-skeleton graphs with $\beta=2$ and different values of the
rectangle side length $a$ and the same number of nodes. In the first
case when $a=1$ the graph corresponds to the classical $\beta$-skeleton
graph in which the nodes are embedded into a unit square. The second
case corresponds to $a=2$ which is a slightly elongated rectangle.

\section{Structural characterization of networks}

One of the goals of this paper is to make a comparison between real-world
fracture networks and the random rectangular neighborhood networks
for different values of the rectangle elongation parameter. This similarity
analysis is carried out by considering a structural characterization
of both types of networks using a series of network-theoretic parameters.
In this section these parameters are defined along with appropriate
references for the reader to dip into their general characteristics.
All these parameters are based on the main matrix representations
of graphs, which are the adjacency $A=\left[A_{ij}\right]_{n\times n}$,
the Laplacian $\mathcal{L}=\left[\mathcal{L}_{ij}\right]_{n\times n}$
and the normalized Laplacian $\hat{\mathcal{L}}=\left[\hat{\mathcal{L}}_{ij}\right]_{n\times n}$
matrices, which are defined respectively as follows:

\begin{center}
$A_{ij}=\left\{ \begin{array}{c}
1\\
0
\end{array}\right.\begin{array}{c}
\textnormal{if }\left(i,j\right)\in E\\
\textnormal{otherwise}
\end{array}\quad\mathcal{L}_{ij}=\left\{ \begin{array}{c}
-1\\
k_{i}\\
0
\end{array}\right.\begin{array}{c}
\textnormal{if }\left(i,j\right)\in E\\
\textnormal{if }i=j\\
\textnormal{otherwise}
\end{array}\quad\hat{\mathcal{L}}_{ij}=\left\{ \begin{array}{c}
-\sqrt{k_{i}k_{j}}\\
1\\
0
\end{array}\right.\begin{array}{c}
\textnormal{if }\left(i,j\right)\in E\\
\textnormal{if }i=j\\
\textnormal{otherwise}
\end{array}$ 
\par\end{center}

There is also the distance matrix $D=\left[d_{ij}\right]_{n\times n}$,
where $d_{ij}$ is the number of edges in the shortest path connecting
$i$ and $j$. The definition of the structural parameters used in
this work are given in Table 1. For the indices of general use see
\cite{Estrada_book} where the indices are explained.

\begin{longtable}{|>{\centering}p{0.3cm}|>{\centering}p{2.5cm}|>{\centering}p{5cm}|>{\centering}p{5.5cm}|>{\centering}p{2.5cm}|}
\caption{Structural parameters used in this work}
\label{table:structural_parameters}\tabularnewline
\hline 
No.  & Index  & Formula  & Observations  & Ref.\tabularnewline
\hline 
\hline 
1  & Average degree  & 
\[
\bar{k}=\dfrac{2m}{n}
\]
 & $m$ is the number of edges in the graph.  & \cite{Estrada_book} \tabularnewline
\hline 
2  & Largest eigenvalue of $A$  & 
\[
\lambda_{1}
\]
 & $\lambda_{1}>\lambda_{2}\geq\cdots\geq\lambda_{n}$, it represents
a sort of average degree in the network.  & \cite{Estrada_book} \tabularnewline
\hline 
3  & Degree variance  & 
\[
V=\dfrac{1}{n}\sum_{i=1}^{n}\left(k_{i}-\bar{k}\right)^{2}
\]
 & A measure of degree irregularity in a graph.  & \cite{degree_variance}\tabularnewline
\hline 
4  & Collatz-Sinogowitz index  & 
\[
\lambda_{1}-\bar{k}
\]
 & A measure of degree irregularity in a graph.  & \cite{Collatz-Sinogowitz}\tabularnewline
\hline 
5  & Degree heterogeneity  & $\rho=\sum_{i,j}\hat{\mathcal{L}}_{ij}$  & A measure of degree irregularity in a graph.  & \cite{degree_heterogeneity}\tabularnewline
\hline 
6  & Degree assortativity  & $r$  & Pearson correlation coefficient of the degree-degree correlation.
$r>0$ (degree assortativity) indicates a tendency of high degree
nodes to connect to other high degree ones. $r<0$ (degree disassortativity)
indicates the tendency of high degree nodes to be connected to low
degree ones.  & \cite{assortativity}\tabularnewline
\hline 
7  & Clustering coefficient  & 
\[
\bar{C}=\dfrac{1}{n}\sum_{i=1}^{n}\dfrac{2t_{i}}{k_{i}\left(k_{i}-1\right)}
\]
 & $t_{i}$ is the number of triangles incident to the vertex $i:$ $t_{i}=\dfrac{1}{2}\left(A^{3}\right)_{ii}$  & \cite{clustering_coefficient}\tabularnewline
\hline 
8  & Diameter  & 
\[
d=\max d\left(i,j\right)
\]
 & The largest entry of the distance matrix $D$.  & \cite{Estrada_book} \tabularnewline
\hline 
9  & Average path length  & 
\[
\bar{l}=\dfrac{1}{n\left(n-1\right)}\sum_{i<j}d\left(i,j\right)
\]
 & A measure of the `small-worldness' of the network.  & \cite{Estrada_book} \tabularnewline
\hline 
10  & Spectral gap  & 
\[
\varDelta=\lambda_{1}-\lambda_{2}
\]
 & A quantity related to isoperimetric properties of graphs. A large
spectral gap indicates the lack of structural bottlenecks in the network.  & \cite{Estrada_book} \tabularnewline
\hline 
11  & Largest eigenvalue of $\mathcal{L}$  & 
\[
\mu_{n}
\]
 & 
\[
0=\mu_{1}<\mu{}_{2}\leq\cdots\leq\mu_{n}
\]
 & \cite{Estrada_book} \tabularnewline
\hline 
12  & Algebraic connectivity  & 
\[
\mu_{2}
\]
 & A measure of the connectivity of the graph.  & \cite{algebraic_connectivity}\tabularnewline
\hline 
13  & Estrada index  & 
\[
EE=\sum_{j=1}^{n}G_{pp}
\]
 & $G_{pp}=\sum_{j=1}^{n}\varphi_{j,p}^{2}\exp\left(\lambda_{j}\right)$
is a centrality measure quantifying the participation of the node
$p$ is subgraphs of the graph, giving more weight to the smaller
ones.  & \cite{Estrada_index}\tabularnewline
\hline 
14  & Spectral bipartivity  & 
\[
b_{s}=\dfrac{\sum_{j=1}^{n}\cosh\left(\lambda_{j}\right)}{\sum_{j=1}^{n}\exp\left(\lambda_{j}\right)}
\]
 & $0<b_{s}\leq1,$ with the lower bound obtained for the complete graph
$K_{n}$ when $n\rightarrow\infty$ and the upper bound obtained for
any bipartite graph.  & \cite{spectral_bipartivity}\tabularnewline
\hline 
15  & Average communicability distance  & 
\[
\bar{\xi}=\dfrac{\sum_{p<q}\left(G_{pp}+G_{qq}-2G_{pq}\right)^{1/2}}{n\left(n-1\right)}
\]
 & A measure of the average quality of communication in a network, where
$G_{pq}=\sum_{j=1}^{n}\varphi_{j,p}\varphi_{j,q}\exp\left(\lambda_{j}\right)$
is the communicability between the corresponding nodes.  & \cite{communicability_distance}\tabularnewline
\hline 
16  & Average communicability angle  & 
\[
\bar{\theta}=\dfrac{\sum_{p<q}\cos^{-1}\left(\dfrac{G_{pq}}{\sqrt{G_{pp}G_{qq}}}\right)}{n\left(n-1\right)}
\]
 & A measure of spatial efficiency of a network. $0\leq\bar{\theta}\leq90$,
where the lower bound indicates high spatial efficiency and the upper
one indicates a poor spatial efficiency.  & \cite{communicability_angle}\tabularnewline
\hline 
17  & Entropy  & $S=-\sum_{j=1}^{n}p_{j}\ln p_{j}$  & A measure of the information content of the spectrum of the adjacency
matrix, where $p_{j}=\dfrac{\exp\left(\lambda_{j}\right)}{EE}$  & \cite{entropy}\tabularnewline
\hline 
18  & Free energy  & $F=-\ln EE$  & A measure of network robustness.  & \cite{entropy}\tabularnewline
\hline 
19  & Kirchhoff index (resistance distance)  & $Kf=\sum_{i<j}\varrho_{ij}$  & A measure related to hitting and commute times in a random walk on
the network, where

$\varrho_{ij}=\sum_{k=2}^{n}\dfrac{1}{\mu_{k}}$ $\left(\psi_{k,i}-\psi_{k,j}\right)^{2}$  & \cite{Kirchhoff_index}\tabularnewline
\hline 
\end{longtable}

The number of 18 small subgraphs are also calculated, labeled as indices
20-37. The formulae for them are found in \cite{Estrada_book} and
a Matlab code for their calculation is provided in Appendix~\ref{appendix:small_subgraphs_code}
of this work.

\section{Properties of RRNGs}

Here we show how some of the properties of the RRNG change with the
elongation parameter, to show why the model can be of practical interest.
We focus on just a few of the most important structural parameters
here: the average node degree, the diameter, and the algebraic connectivity
$\mu_{2}$. First we prove the following result about the diameter
of the RRNG.
\begin{lemma}
\label{thm:diameter}Let $G=G\left(n,a,\beta=2\right)$ be a connected
RRNG with $n$ nodes embedded in a rectangle of sides with lengths
$a$ and $b=a^{-1}$. Let $D=D\left(G\right)$ be the diameter of
the corresponding RRNG. Then, 
\end{lemma}
\begin{equation}
D\left(G\right)\geq\dfrac{\sqrt{\left(a^{4}+1\right)n}}{a}.\label{eq:diameter}
\end{equation}

\begin{svmultproof}
The nodes of the RRNG are uniformly and independently distributed
in the unit rectangle. Then, let us assume that the $n$ points are
equally spaced in the area of the rectangle separated by a Euclidean
distance $L$. In this case the largest number of points are connected
along the main diagonal of the rectangle. If the length of the main
diagonal is $c$ there are $\dfrac{c}{L}$ connected nodes in this
line. Thus, the maximum shortest path distance in the RRNG is $\dfrac{c}{L}$
with $c=\dfrac{\sqrt{a^{4}+1}}{a}$. For a connected RNG this is the
shortest the diameter can be, because if two points in the main diagonal
are separated at a Euclidean distance larger than $L$, then the diameter
of $G$ will be larger than $\dfrac{c}{L}$. Now, the main problem
here is to determine the value of $L$ for the separation of two points
in the RRNG. Here we consider that the $n$ points can be distributed
in the square ($a=1$) forming a regular square lattice. In this case
there should be $\sqrt{n}$ rows of $\sqrt{n}$ points equally spaced
in the square. Thus, the separation between two points is $L=1/\sqrt{n}$.
In the case of the rectangle we follow a naive approach of considering
that a rectangle of major length $a>1$ can be obtained just by cutting
the square $a$ times in the direction of the $y$-axis and pasting
the cut rectangle along the $x$-axis. In this way it is guarantee
that the separation between two points in the original square remains
the same in the elongated rectangle. That is, $L=1/\sqrt{n}$ for
any value of $a$, such that we have the final result.
\end{svmultproof}

\begin{figure}[H]
\begin{centering}
\includegraphics[width=0.33\textwidth]{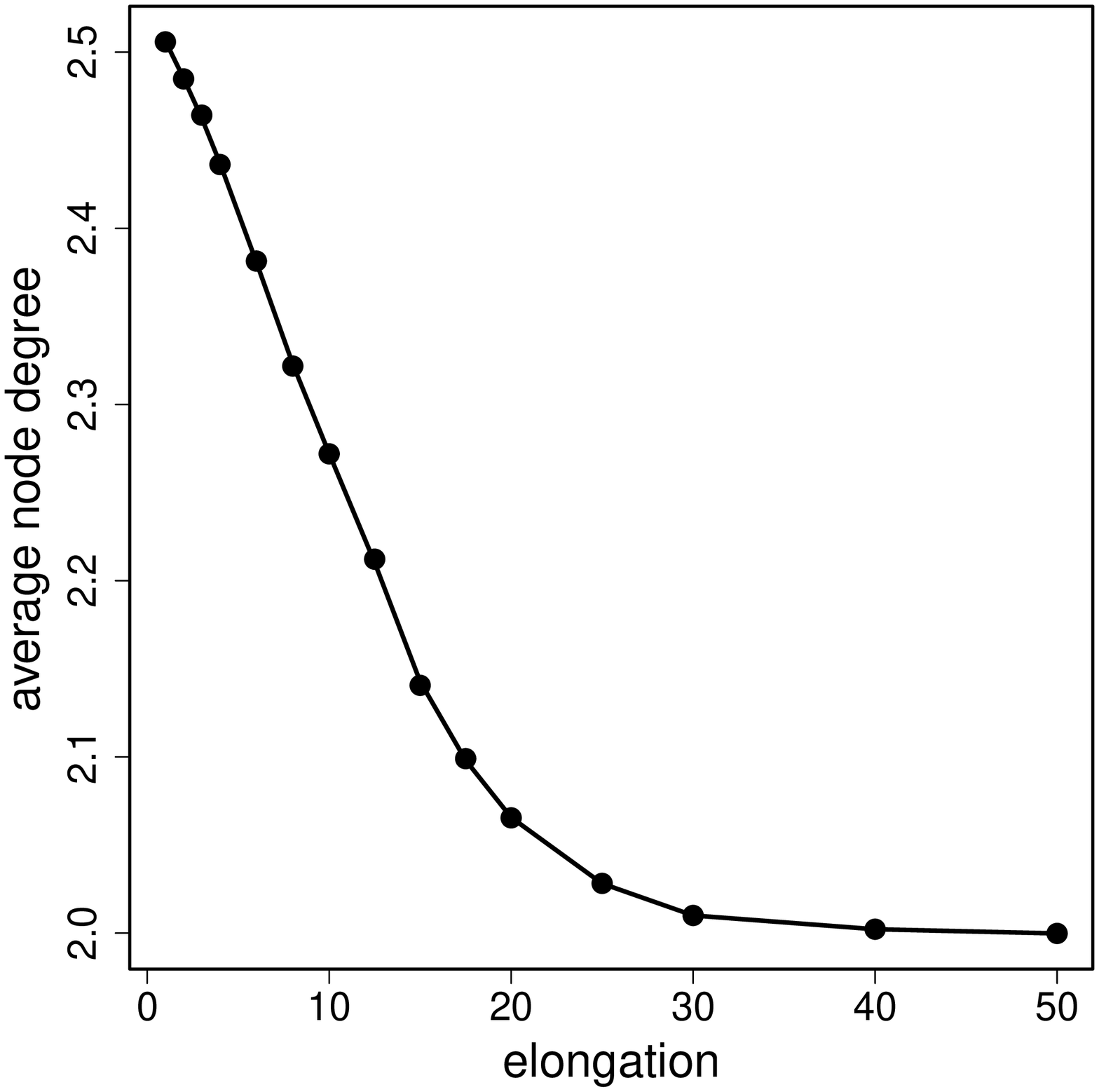}\includegraphics[width=0.33\textwidth]{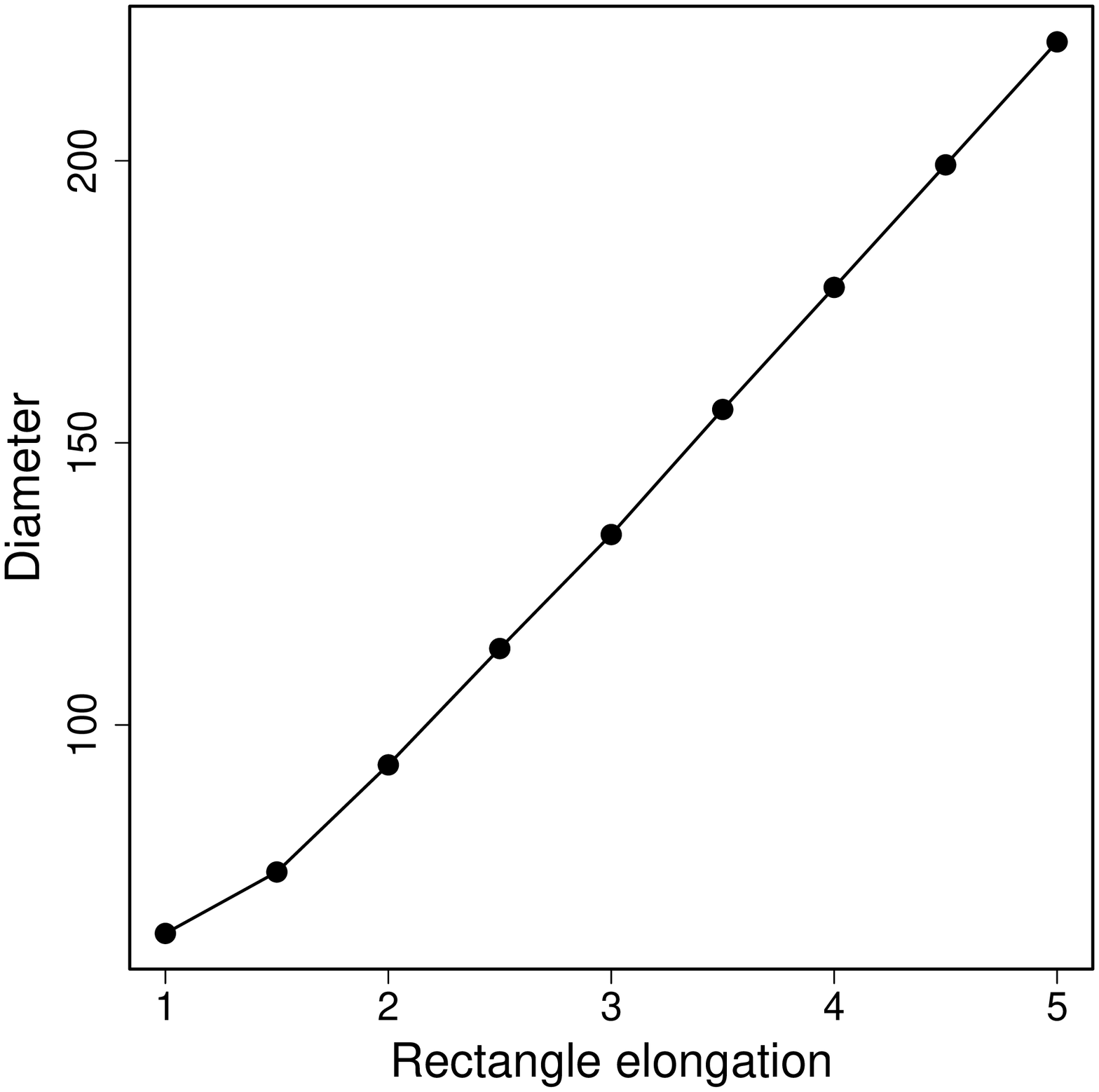} 
\par\end{centering}
\begin{centering}
\includegraphics[width=0.33\textwidth]{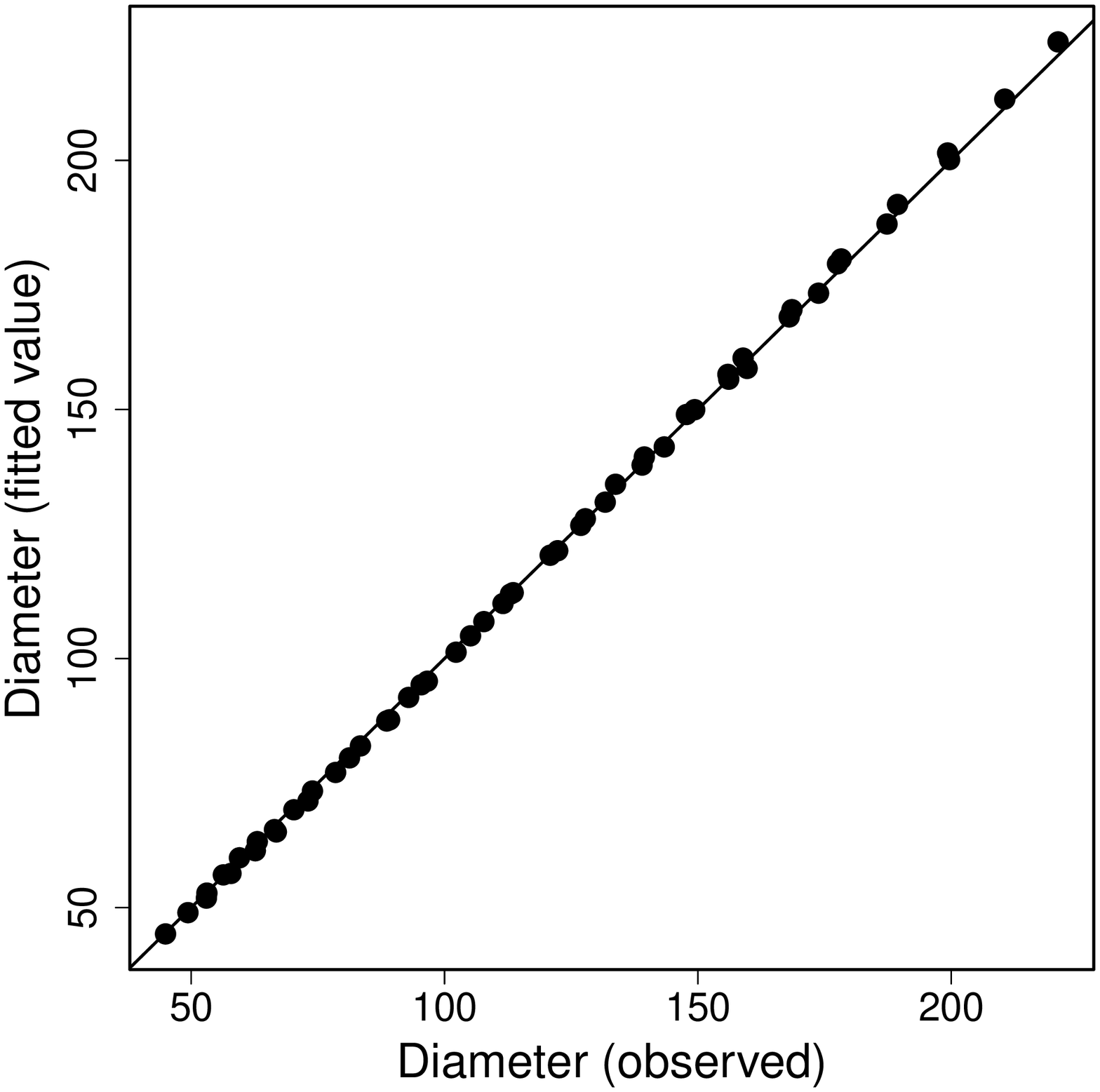}\includegraphics[width=0.33\textwidth]{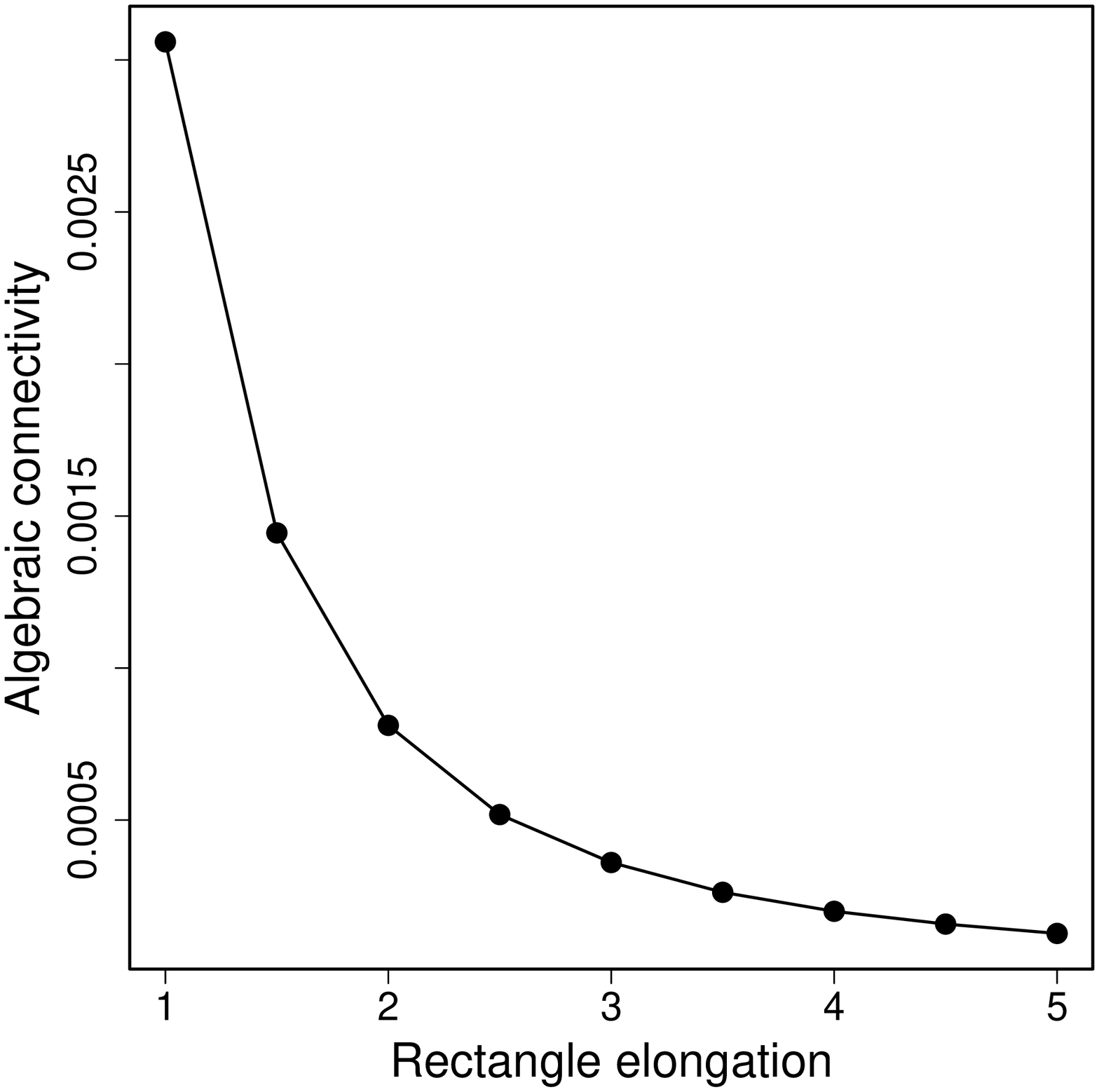} 
\par\end{centering}
\caption{Plots of how several structural parameters of the RRNG model changes
as the elongation of the rectangle varies from $a=1$ to $a=5$ for
$n=1,000$ nodes: (a) average node degree, (b) diameter, (c) $\mu_{2}$.}

\label{f;rng_properties} 
\end{figure}

Obviously, we can fit the values of the right-hand-side of (\ref{eq:diameter})
to the actual values of the diameter in RRNGs of different sizes.
That is, we can obtain an empirical relation of the kind $D\left(G\right)\approx c\dfrac{\sqrt{\left(a^{4}+1\right)n}}{a},$
where $c$ is a fitting parameter. By doing so for RRNGs with sizes
$n=500,...,1000$ we found that $c\approx1.414$ and the correlation
coefficient between the observed and the calculated diameter is larger
than 0.999 (see the plot in Figure \ref{f;rng_properties}).

The previous results is very important because it allow us to bound
the algebraic connectivity of the RRNG. The algebraic connectivity\textendash as
we will see later on this paper\textendash is a fundamental parameter
for understanding the diffusive processes taking place on RRNGs. Then,
we prove here the following result.
\begin{lemma}
Let $G=G\left(n,a,\beta=2\right)$ be a connected RRNG with $n$ nodes
embedded in a rectangle of sides with lengths $a$ and $b=a^{-1}$.
Let $k_{max}$ be the maximum degree of any node in $G$. Then, the
second smallest eigenvalue of the Laplacian matrix of the RRNG is
bounded as 
\end{lemma}
\begin{equation}
\mu_{2}\left(G\right)\leq\dfrac{8k_{max}a^{2}}{a^{4}+1}\dfrac{\log_{2}^{2}n}{n},\label{eq:final bound}
\end{equation}

\begin{svmultproof}
We simply substitute our bound for the diameter of the RRGN into the
Alon-Milman bound for the algebraic connectivity of any graph \ref{eq:Alon Milman-1},

\begin{equation}
\mu_{2}\left(G\right)\leq\dfrac{8k_{max}}{D^{2}}\log_{2}^{2}n.\label{eq:Alon Milman-1}
\end{equation}
\end{svmultproof}

It should be noticed that in RRNGs the value of $k_{max}$ is typically
quite small. In all our simulations it is not bigger than $6$. These
two theoretical results clearly indicate that the properties of the
RRNG are significantly and non-trivially affected by the elongation
of the rectangle. The diameter of the RRNGs increases almost linearly
with the elongation of the rectangle. On the other hand, the elongation
of the rectangle in the RRNG makes the graphs drastically less connected.
In order to see these effects in practice we develop a series of simulations
considering RRNGs with $n=1000$ nodes, $1\leq a\leq5$ in steps of
$0.5$, which we find to be a sufficiently large range of elongations
for the analyses in this work. It is clear that in the limit as $a\rightarrow\infty$
the RRNG tends to the path graph $P_{n}$, which will dictate the
behavior at larger elongations not seen in these plots, except for
the average node degree to demonstrate this behavior. As the elongation
increases, the average node degree decreases since the perimeter increases
and nodes near the boundary are surrounded by fewer nodes to attach
to; with further elongation the value will approach $\bar{k}=2-2/n\simeq2$.
As predicted by our theoretical results, the diameter of the RRNGs
increases almost linearly with the elongation of the rectangle and
the algebraic connectivity decay in a nonlinear fashion with it. Therefore,
we conclude that as the elongation of the RRNG model is increased,
the structural parameters also change, including those closely connected
to dynamical processes such as the algebraic connectivity.

\section{Rock Fracture Networks}

\begin{figure}[H]
\begin{centering}
\includegraphics[bb=0bp 0bp 486bp 315bp,clip,width=0.75\textwidth]{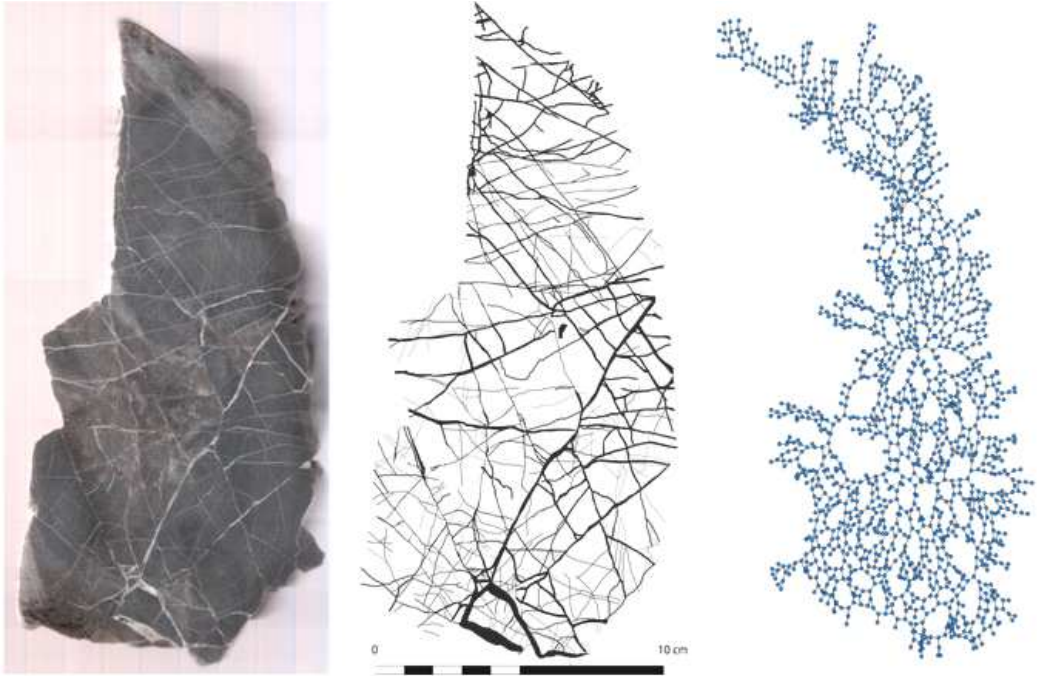} 
\par\end{centering}
\caption{Illustration of the process of creating rock fracture networks from
rock samples. (Left) A rock sample from the Gulf of Mexico showing
one of the halves of the rock. (center) Digital image of the rock
illustrating the fractures existing in it. (right) Rock fracture network
created from the digital image by taking the intersection of channels
as the nodes and channels as the edges of the network. Image is courtesy
of E.~Santiago.}

\label{Rock images} 
\end{figure}

This section describes the dataset of real-world networks consisting
of the channels and their intersections produced by fractures in rocks
of petrophysical interest. The procedures described hereafter are
based on the analysis developed by \cite{Santiago_1}. These authors
have considered a series of rocks extracted from wells in the Gulf
of Mexico. The rocks are cut into two halves and images are taken
of one of the rock halves, which show the fractures in the corresponding
rock. An algorithm is then used to find the skeleton of these fractures
and construct a network representation of it, which is stored as an
adjacency matrix. The nodes of the network correspond to where fractures
intersect or terminate and an edge between nodes corresponds to a
channel between those points in the skeleton of the rock fracture.
A sketch of the process is illustrated in Fig.~\ref{Rock images}.

In total 29 rock samples are considered here, kindly provided by \cite{Santiago_1,Santiago_metrics,Santiago_centralities}.
The number of nodes $n$ and edges $m$ in the 29 networks studied
here are provided in Table (\ref{Rock labels}) together with the
labels used in the subsequent analysis in this work.

\begin{table}[H]
\begin{centering}
\begin{tabular}{|c|>{\centering}p{2cm}|>{\centering}p{2cm}|>{\centering}p{2cm}|>{\centering}p{0.5cm}|>{\centering}p{1cm}|>{\centering}p{2cm}|>{\centering}p{2cm}|>{\centering}p{2cm}|}
\hline 
No.  & $n$  & $m$  & $a$ &  & No.  & $n$  & $m$ & $a$\tabularnewline
\hline 
\hline 
M1  & 93  & 109  & $1.0$ &  & M16  & 779  & 887 & $1.5$\tabularnewline
\hline 
M2  & 46  & 54  & $1.5$ &  & M17  & 633  & 728 & $1.5$\tabularnewline
\hline 
M3  & 71  & 74  & $2.0$ &  & M18  & 3585  & 4183 & $1.5$\tabularnewline
\hline 
M4  & 109  & 124  & $1.5$ &  & M19  & 97  & 101 & $2.0$\tabularnewline
\hline 
M5  & 346  & 380  & $3.5$ &  & M20  & 1567  & 1996 & $1.0$\tabularnewline
\hline 
M6  & 85  & 89  & $2.5$ &  & M21  & 70  & 74 & $2.0$\tabularnewline
\hline 
M7  & 55  & 60  & $1.0$ &  & M22  & 1686  & 1905 & $2.0$\tabularnewline
\hline 
M8  & 87  & 94  & $1.0$ &  & M23  & 396  & 396 & $4.5$\tabularnewline
\hline 
M9  & 44  & 46  & $1.0$ &  & M24  & 181  & 180 & $5.0$\tabularnewline
\hline 
M10  & 296  & 336  & $1.5$ &  & M25  & 394  & 418 & $4.0$\tabularnewline
\hline 
M11  & 47  & 48  & $2.0$ &  & M26  & 808  & 813 & $4.5$\tabularnewline
\hline 
M12  & 46  & 47  & $1.5$ &  & M27  & 223  & 222 & $4.0$\tabularnewline
\hline 
M13  & 215  & 233  & $2.5$ &  & M28  & 297  & 305 & $4.0$\tabularnewline
\hline 
M14  & 132  & 144  & $1.5$ &  & M29  & 363  & 365 & $5.0$\tabularnewline
\hline 
M15  & 40  & 40  & $2.0$ &  &  &  &  & \tabularnewline
\hline 
\end{tabular}
\par\end{centering}
\caption{Rock fracture networks studied in this work, their number of nodes
$n$, and number of edges $m$.}

\label{Rock labels} 
\end{table}

\section{Similarity between Fracture Networks and RNGs}

This section aims to compare the real-world rock fracture networks
with their random analogues created by using the $\beta$-skeleton
approach. To achieve this goal random rectangular neighborhood graphs
are created with a value of $\beta=2$ and having the same number
of nodes and edges as the corresponding real-world fracture network.
The elongations of the rectangle are varied in the range $1\leq a\leq5$
with a step $0.5$. The selection of the value $\beta=2$ in this
study is based on empirical observations of the dataset of rock fracture
networks under analysis. First, these real-world networks are always
planar and have a relatively low number of triangles. These two characteristics
are well-reproduced by the relative neighborhood graphs corresponding
to $\beta=2$. Furthermore, these RNGs have been widely considered
in the literature and can be constructed more easily and quickly then
an a $\beta$-skeleton for some arbitrary value of $\beta$.

Then, for each graph, a $k\times1$ vector is calculated consisting
of the $k$ structural properties defined previously (including the
18 small subgraphs). That is, every network is represented in a $k$-dimensional
space ($k=37)$ in which each coordinate represents a structural parameter,
e.g., average degree, clustering coefficient, etc. A number of random
constructions are realized of the RRNG for each elongation and the
value averaged over all of them. The number of random realizations
varies depending on the size of the network for reasons of computational
difficulty, with effort made to ensure a large enough number of realizations
such that the variance is not too large.

The similarity between the real-world rock fracture networks and their
RRNG analogues can then be calculated, we may also refer to the dissimilarity
which is simply a different view of the same quantity. This similarity
is quantified by simply using the Euclidean distance between the corresponding
points in the $k$-dimensional property space in which they are represented,
and thus the similarity may referred to as the ``distance'' when
convenient. A potential problem arising here is the fact that the
values of the properties calculated lie in a very wide range of numerical
values. Then, the values of each property are normalized to lie in
the range between 0 and 1. Such normalization is carried out as follows.
Let $r_{i,0}$ be the property vector for the $i$th RFN, and $r_{i,j}$
be the property vector for the RRNG analogous of the corresponding
RFN created with the $j$th elongation $a$ of the rectangle, $1\leq j\leq9$,
and for each elongation this is averaged over all random realizations
to obtain a single vector for each. Each vector $r_{i,j}$ is then
normalized as follows for $0\leq j\leq9$

\begin{equation}
\hat{r}_{i,j}(p)=\dfrac{r_{i,j}\left(p\right)-\min_{j}r_{i,j}\left(p\right)}{\max_{j}r_{i,j}\left(p\right)-\min_{j}r_{i,j}\left(p\right)},\label{eq:normalisation}
\end{equation}

where $r_{i,j}\left(p\right)$ represents the $p$th entry of the
$r_{i,j}$ vector. That is, for a given RFN and property, the value
of this property is normalized for the RFN and all corresponding elongations
of RRNG so that they lie between 0 and 1, with the smallest value
mapped to 0 and the largest mapped to 1.

\begin{figure}[H]
\begin{centering}
\includegraphics[width=0.45\textwidth]{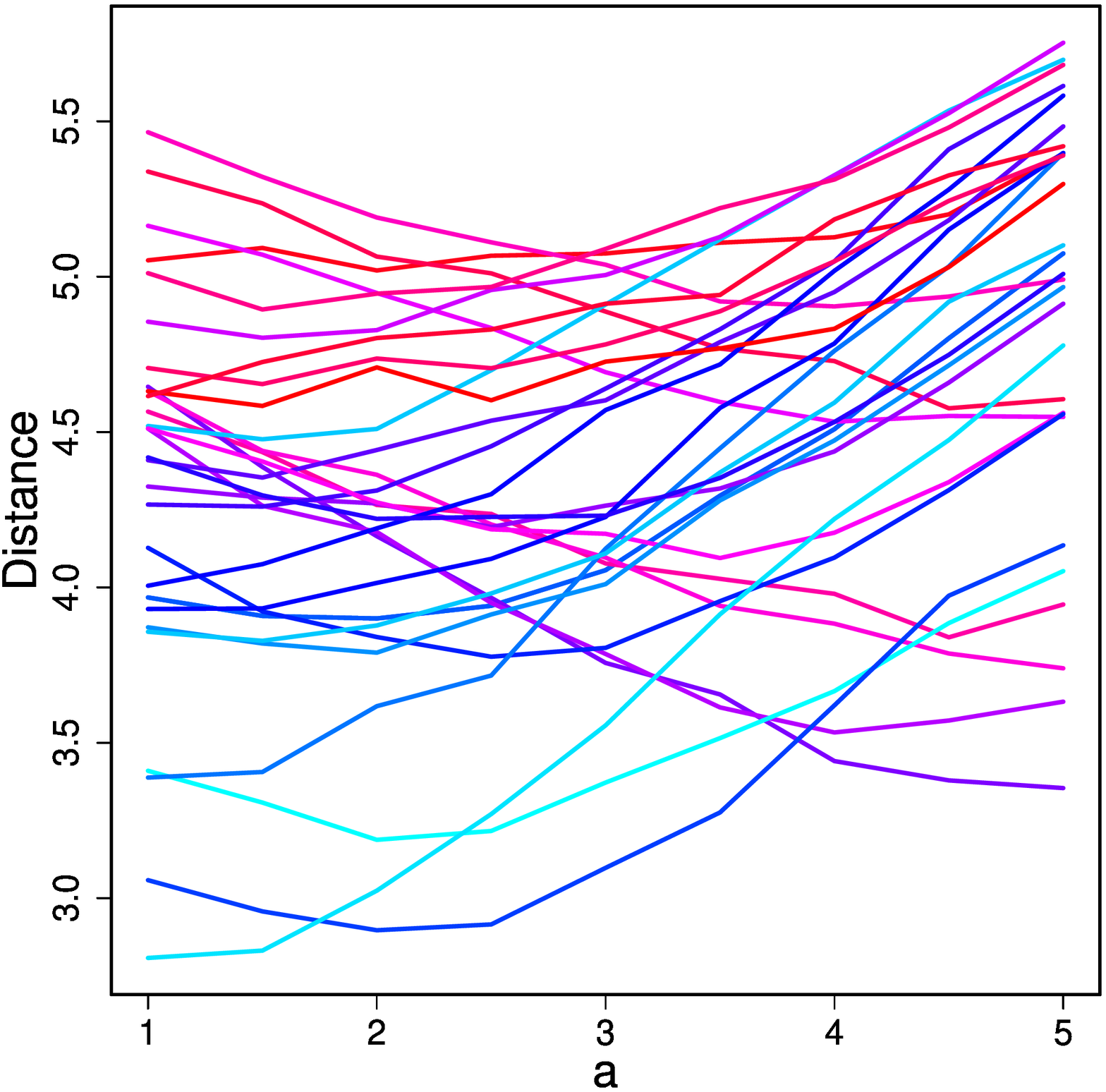}\includegraphics[width=0.45\textwidth]{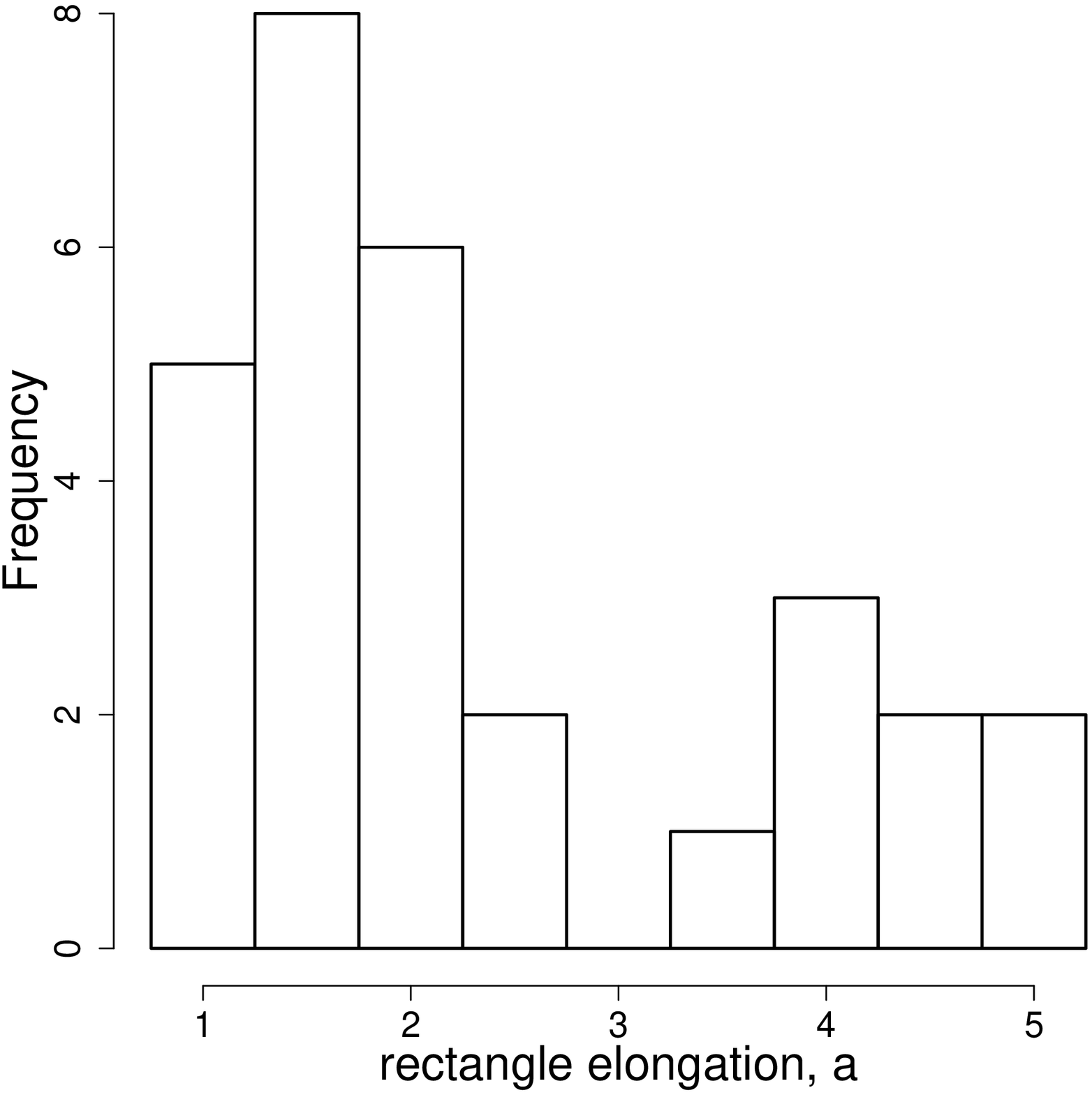}
\includegraphics[width=0.45\textwidth]{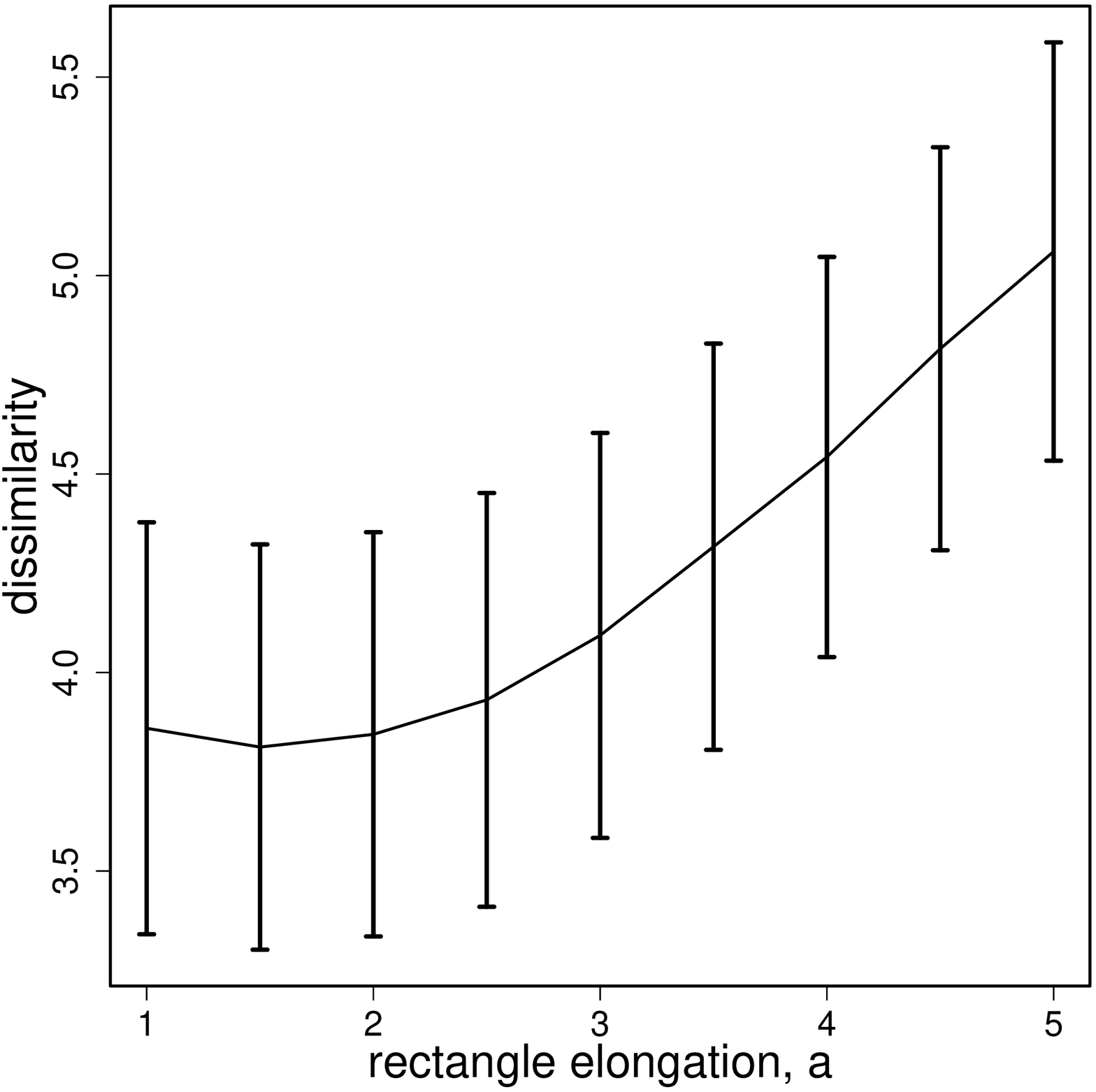}\includegraphics[width=0.45\textwidth]{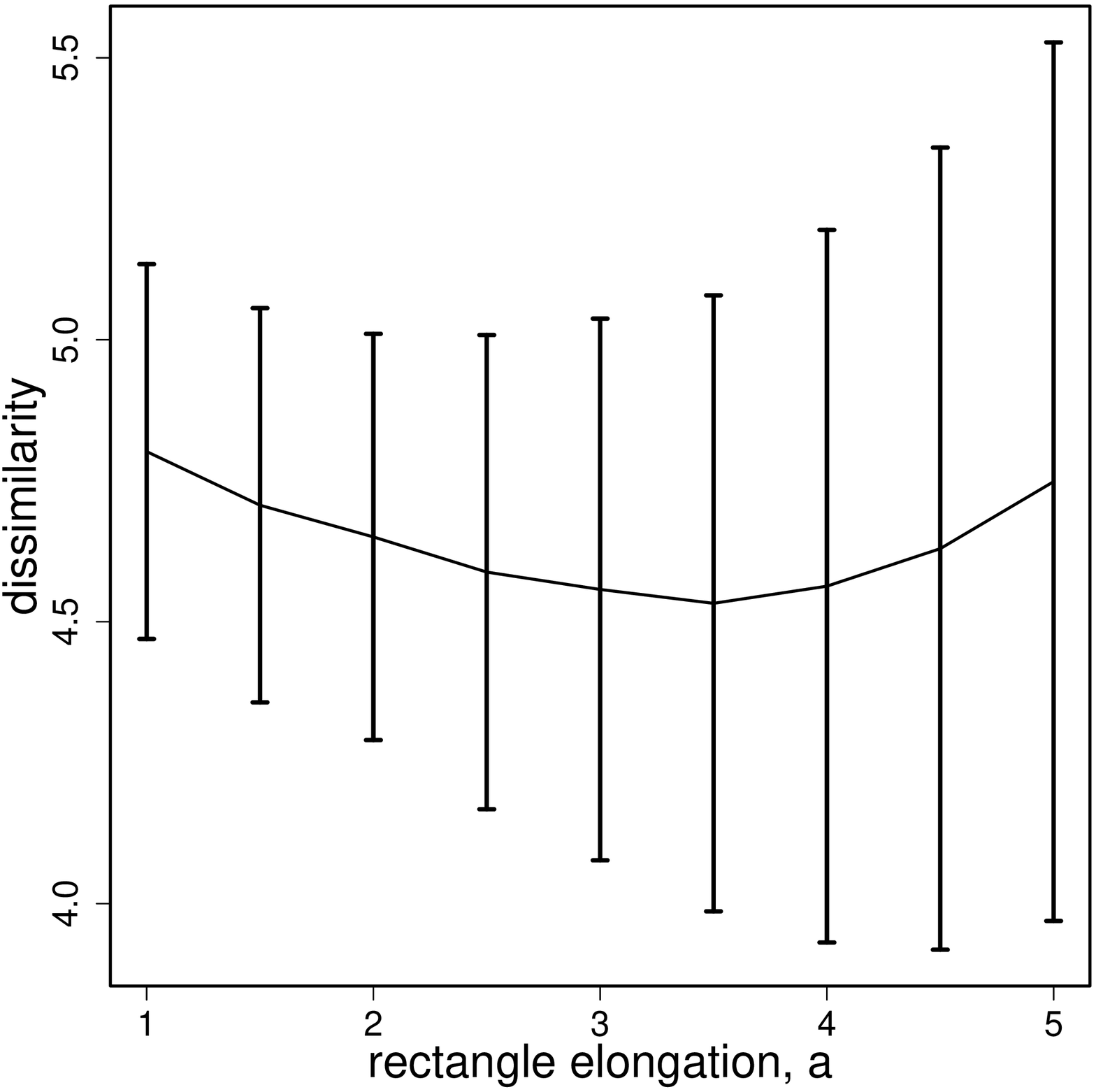} 
\par\end{centering}
\caption{(a) Variation of the dissimilarity between RFNs and the corresponding
RRNGs for different values of the rectangular elongation $a$. Each
curve corresponds to one pair RFN-RRNG. (b) Number of graphs for which
the minimum dissimilarity appears at a given value of the rectangle
elongation $a$. (c) Average variation of the dissimilarity between
RFNs and the corresponding RRNGs for different values of the rectangular
elongation $a$ for RFNs with less than 150 nodes. The vertical bars
represent the standard deviation. (d) Average variation of the dissimilarity
between RFNs and the corresponding RRNGs for different values of the
rectangular elongation $a$ for RFNs with more than 150 nodes. The
vertical bars represent the standard deviation. }

\label{similarity plots} 
\end{figure}

It is observed that for each of the rock fracture networks there is
a minimum in the plot of the dissimilarity versus the rectangle elongation
(see Fig.~\ref{similarity plots}(a)), which indicates that there
is an optimal elongation for each RRNG that makes it most similar
to the real-world fracture network. In Fig.~\ref{similarity plots}(b),
the frequency with which the maximum similarity occurs for a given
value of the rectangle elongation $a$ is plotted. It can be seen
that the histogram is two-peaked with the first maximum corresponding
to elongations between $a=1$ and $a=2$ and the second one for elongations
around $a=4$. The first peak clearly corresponds to rock fracture
networks that are better reproduced by almost square neighborhood
graphs. However, the second group of real-world fracture networks
are better reproduced by elongated rectangles in which one of the
sides of the rectangle is about 16 times longer than the other. A
more detailed statistical analysis of the histogram of the optimal
elongations shows that the modal value of elongation is $a=1.5$,
and elongations between $a=1$ and $a=2$ are the most common, accounting
for about $2/3$ of the rocks. The rest of the values lie between
$a=2.5$ and $a=4.5$, except for a couple of rocks which have minimum
at $a=5.$

The most interesting observation carried out in this analysis is the
following. Most of the rock fracture networks which are better reproduced
by almost square RRNGs correspond to the smallest ones, while those
which are better reproduced by elongated RRNGs are those having the
largest number of nodes. These results are illustrated in Fig.~\ref{similarity plots}(c
and d) where the networks are split into two groups, those with $n\leq150$,
and those with $n>150$. As can be seen in Fig.~\ref{similarity plots}(c),
which corresponds to the first group, the maximum similarity occurs
for $1\leq a\leq2$. However, for the largest networks the maximum
similarity occurs for values of $3\leq a\leq4$. In closing, the rock
fracture networks are better described by the RRNGs depending on the
size of the networks, with almost square RRNG describing better the
smallest RFNs and more elongated RRNGs describing better the largest
RFNs. In other words, it is more plausible that larger RFNs are those
coming from elongated rocks and consequently better reproduced by
RRNGs with $a>1$ which better reproduce this characteristic. The
smallest RFNs appear to come from more rounded rocks, which are better
reproduced by almost-square RRNGs.

The RRNGs reproduce relatively well the main structural properties
of real-world RFNs of different sizes. This conclusion is reached
by the fact that the dissimilarity between the RFNs and the RRNGs
is generally smaller than the dissimilarity between the RFNs and other
commonly used random graph models. A comparison is made here between
the use of the RRNGs and the Erd\H{o}s\textendash Rényi (ER), Barabási\textendash Albert
(BA), and random rectangular graph (RRG) models. The ER graphs $G\left(n,p\right)$
are created from a set of $n$ nodes which are randomly picked in
pairs and connected with certain probability $p$. In the BA model
a graph $G\left(n,m_{0}\right)$ is created on the basis of a seed
of $n_{0}$ nodes which are connected randomly and independently according
to an ER model. Then, new nodes are added one at a time and connected
to $m_{0}$ of the existing nodes in the graph with a probability
that depends on the degree of the nodes. Then, instead of obtaining
a Poisson distribution of the node degree as in the ER model, the
model ends up with a power-law degree distribution. Finally, the RRG
$G\left(n,a,r\right)$ is created from $n$ points randomly and independently
distributed on the unit rectangle of sides $a$ and $1/a$. Then,
a disk of radius $r$ is centered on a point $i$, and every node
inside that disk is connected to $i$. The process is repeated for
each of the $n$ points. Notice that all the random graphs must be
connected in order to calculate some of the structural parameters.
For the ER model the value of $p=\log n/n$ was chosen, which is the
critical value at which the large connected component appears, with
discarding of any random realizations in which the networks were disconnected.
For the RRG model, again elongations from $a=1$ to $a=5$ in steps
of $0.5$ were used, and the best elongations were selected for each
rock individually in exactly the same manner as for the RRNG model.
In every case, the radius $r$ was selected to ensure the RRG was
connected with high probability, since selecting $r$ to best match
the number of edges in the RFN would almost always give a disconnected
graph. In each case, several random realizations were taken and the
results averaged, with fewer realizations used in the case of a large
number of nodes for reasons of computational difficulty.

\begin{figure}[H]
\begin{centering}
\includegraphics[width=0.33\textwidth]{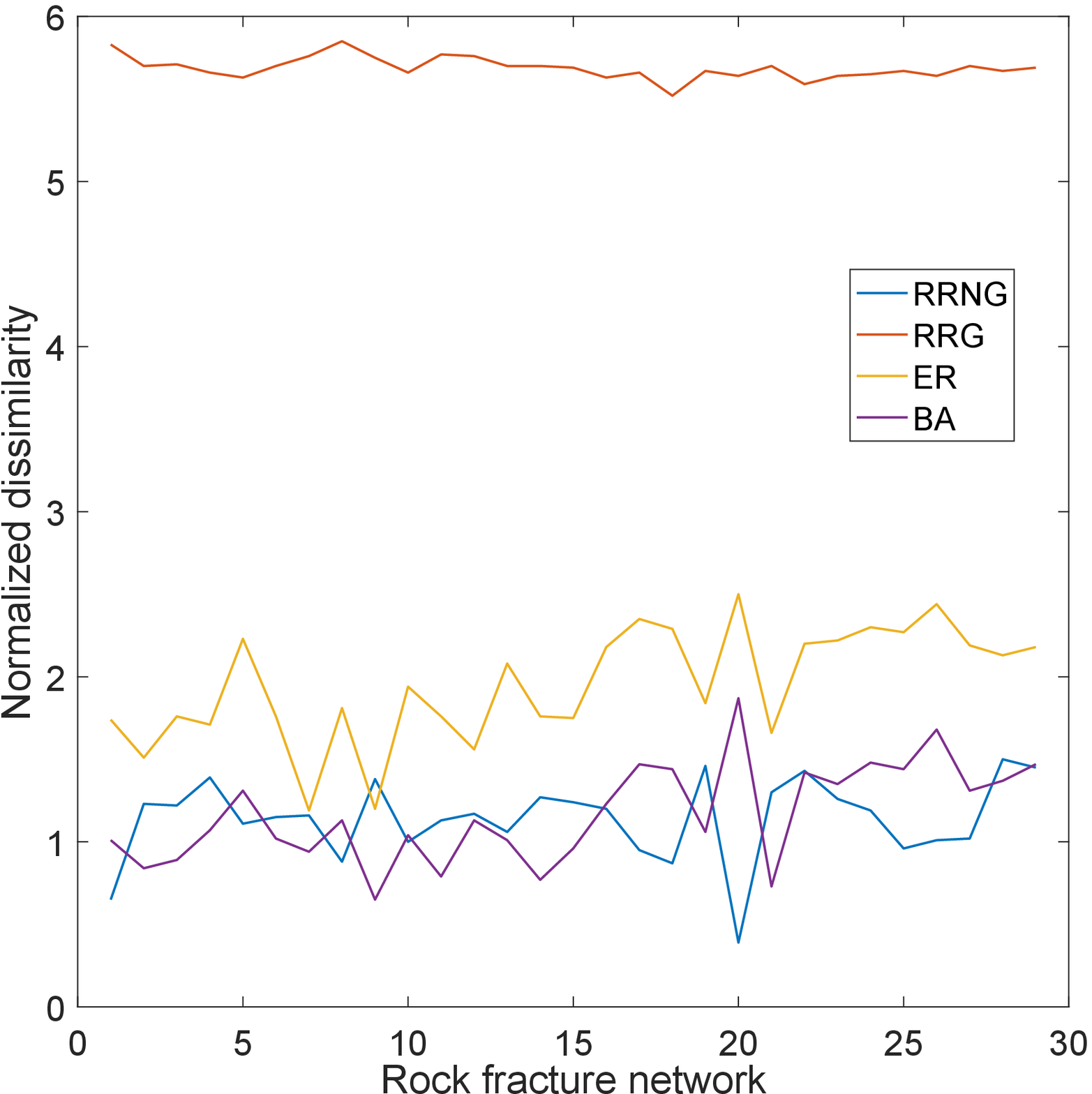}\includegraphics[width=0.33\textwidth]{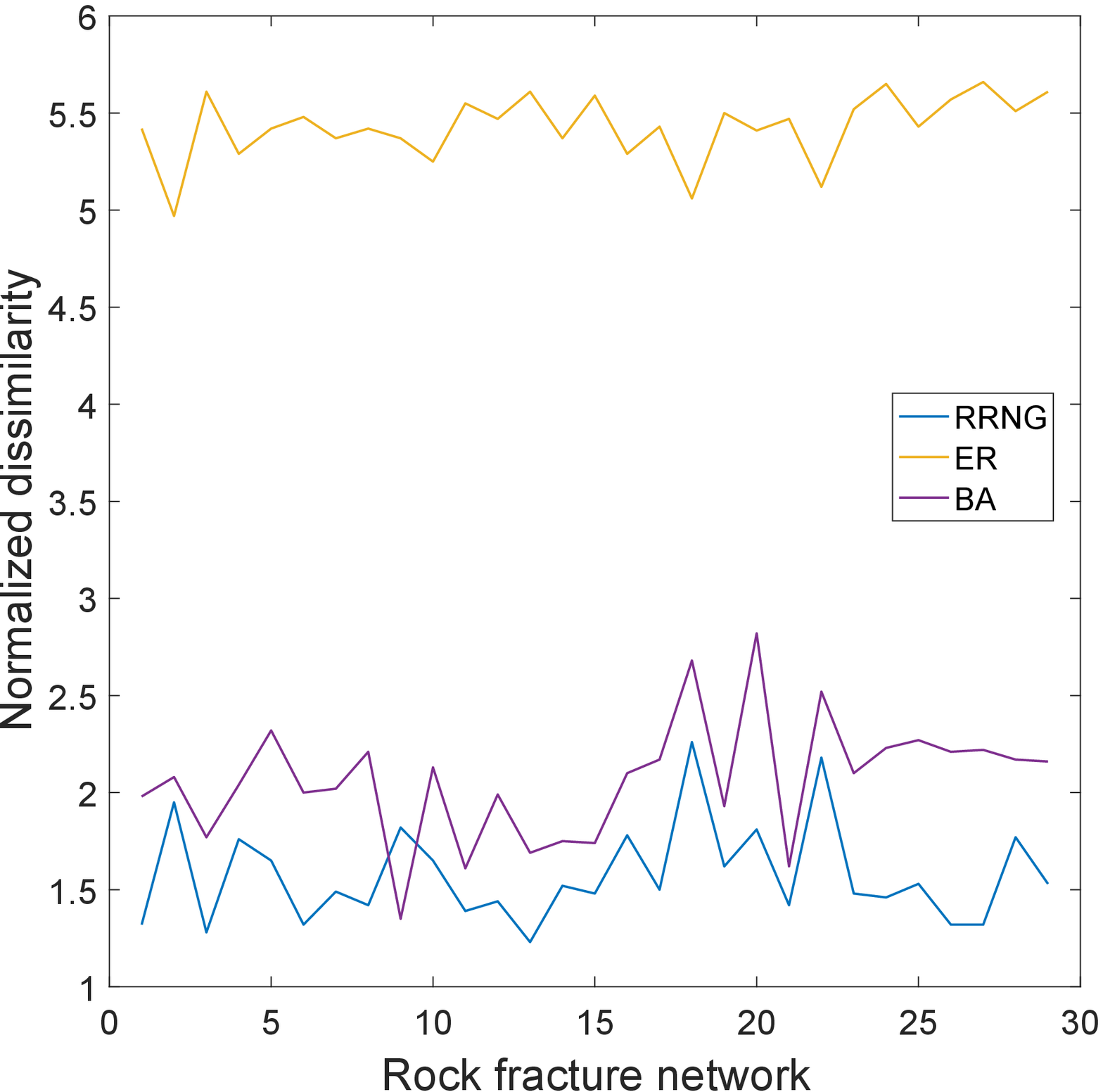}\includegraphics[width=0.33\textwidth]{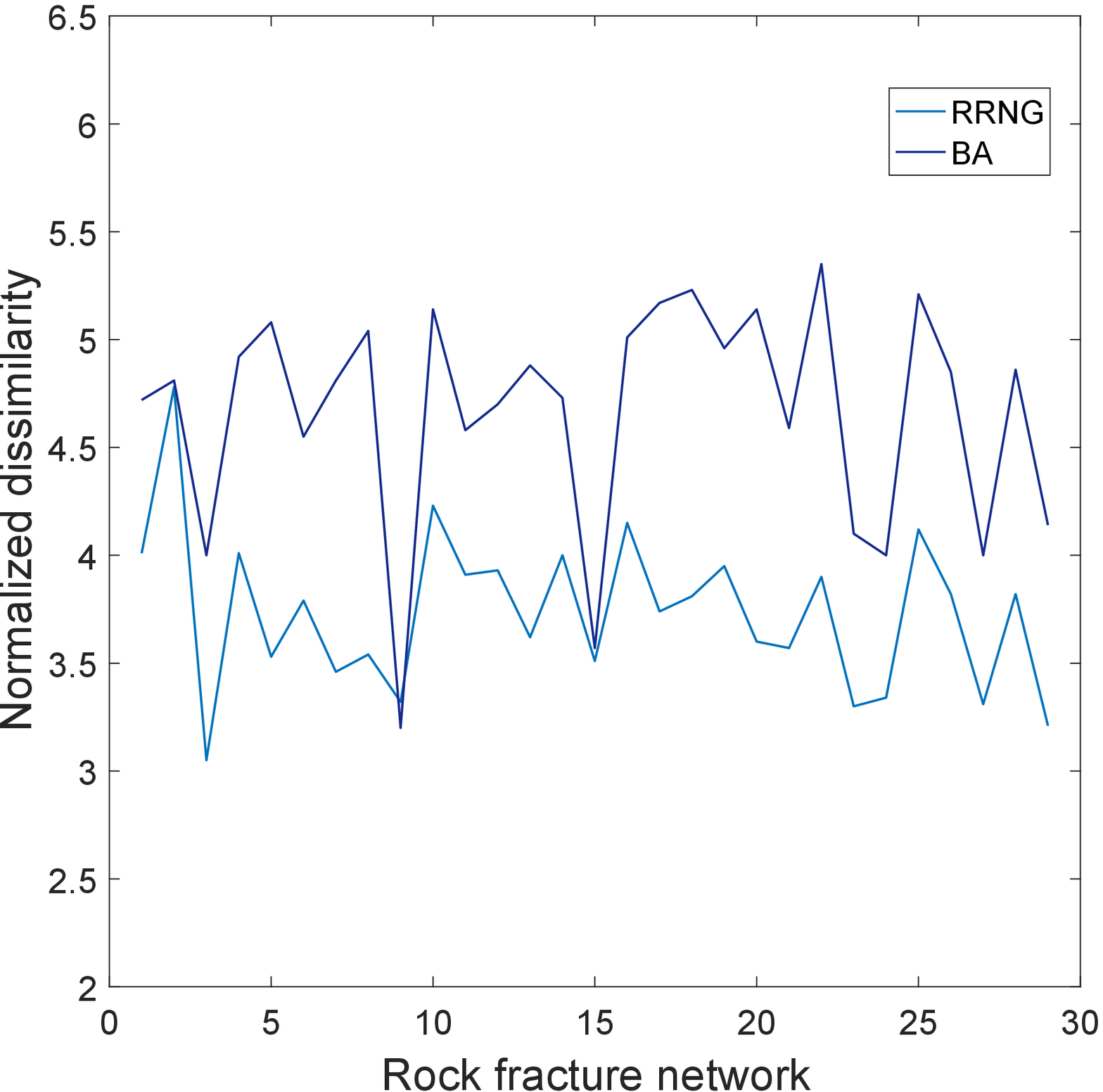} 
\par\end{centering}
\caption{Relative normalized dissimilarity (the dissimilarity after the data
has been normalized as described in the main text) between the RFNs
and four theoretical models: random rectangular neighborhood networks
(RRNG), random rectangular graphs (RRG), Erd\H{o}s-Rényi (ER) and
Barabási-Albert (BA). From left to right the iterative process described
in the main text is shown where the least similar model is eliminated
until the two best models remain.}

\label{Similarity comparison} 
\end{figure}

With this data, the models were compared to see which of them works
the best. Let $r_{i,0}$ be the property vector for the $i$th RFN,
and $r_{i,j}$ be the property vector for the $j$th model, $1\leq j\leq4$,
where for the RRNG and the RRG only the best elongation is used. Then,
the normalization takes exactly the same form as Eq.~\ref{eq:normalisation}.
After this normalization, dissimilarities between the RFNs and each
model were calculated. As can be seen in Fig.~\ref{Similarity comparison}(a)
the RRG model is the worst, with a dissimilarity larger than 5 for
all RFNs, which is more than double the largest dissimilarity for
any other model. To better compare the remaining models, the RRG model
was removed from the analysis and the dissimilarities were recomputed
with the remaining three models, since the presence of the data from
the RRG model affects the normalization values. The results are illustrated
in Fig.~\ref{Similarity comparison}(b) where it can be noted that
the ER model is much worse than the other two, so this model was also
removed. The result is the comparison of the best two models, the
RRNG and the BA, which is illustrated in Fig.~\ref{Similarity comparison}(c).
In the large majority of cases the RRNG is observed to be the best
model to reproduce the topological properties of the rock fracture
networks due to a smaller value of dissimilarity.

\section{Fluid diffusion on fracture networks and on RRNGs }

In this section a diffusion process is modeled as occurring through
the channels of rock fracture networks. The flow of fluids through
porous media is frequently described by the Boussinesq equation (also
known as the porous media equation):

\begin{eqnarray}
\dfrac{\partial}{\partial t}u\left(x,t\right) & = & \dfrac{\partial}{\partial x}\left(u\left(x,t\right)\dfrac{\partial}{\partial x}u\left(x,t\right)\right),\\
u\left(x,0\right) & = & u_{0}
\end{eqnarray}

where $u\left(x,t\right)$ is a non-negative scalar function on $x\in\Omega=\left[0,1\right]\subset\mathbb{R}$
and time $t\in\mathbb{R}\geq0$.

Suppose that the fluid is flowing through a capillary of length $L$
and height $h_{0}$, and suppose that the capillary is much longer
than it is thick: $L\gg h_{0}.$ Then, according to \cite{porous_media_1}
(see especially Fig.\ 4) and \cite{porous_media_2}, the Boussinesq
equation can be very well approximated by a simple heat equation

\begin{eqnarray}
\dfrac{\partial}{\partial t}u\left(x,t\right) & = & \dfrac{\partial^{2}}{\partial x^{2}}u\left(x,t\right),\\
u\left(x,0\right) & = & u_{0}.
\end{eqnarray}

Obviously, the channels produced by the fractures of rocks are less
than a millimeter thick and a few centimeters long. Thus, $L\gg h_{0}$
is always true and the use of the heat equation is justified for modeling
the diffusion of oil and gas through the channels formed by the network
of rock fractures. In the case of diffusion through the edges of a
network the previous equation can be written as

\begin{eqnarray}
\dfrac{\partial}{\partial t}u\left(x,t\right) & = & -\mathcal{L}u\left(x,t\right),\label{eq:diffusion equation}\\
u\left(x,0\right) & = & u_{0},\label{eq:initial condition}
\end{eqnarray}
where $\mathcal{L}$ is the graph Laplacian defined previously in
this work.

\begin{figure}[H]
\begin{centering}
\includegraphics[width=0.75\textwidth]{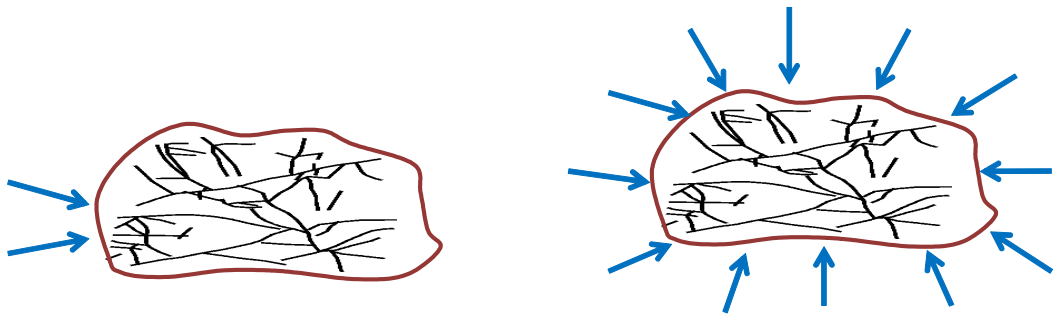} 
\par\end{centering}
\caption{Illustration of the two scenarios used for modeling the diffusion
of oil and gas from a reservoir through the fracture network of a
rock. (a) Only a small region of the rock is in contact with the reservoir.
(b) The whole rock is in contact with the reservoir. }

\label{initial conditions} 
\end{figure}

In order to select the initial condition vector $u_{0}$ two different
scenarios are considered. The first one considers the case in which
only a few nodes of the rock fracture network are in contact with
the reservoir as illustrated in Fig.~\ref{initial conditions}(a).
This initial condition is used mainly to study the influence of the
flow directionality on the diffusion through the rock fracture network
and will be referred to as directional diffusion. In this case the
vector $u_{0}$ is constructed such that the entry $u_{0}\left(i\right)$
is a random number for those nodes considered to be in contact with
the reservoir, which are randomly selected from the set of nodes of
the graph and should have the condition that they are close in space
to each other. This set of points is selected from one of the two
extremes of the longest path (diameter) of the graph. The rest of
the entries of the initial condition vector are set to zero. The second
scenario is based on the assumption that the RFN is in contact with
the reservoir from many different positions as illustrated in Fig.~\ref{initial conditions}(b),
and will be called anisotropic diffusion. In this case the entries
$u_{0}\left(i\right)$ of the initial condition vector are selected
randomly for each of the nodes of the graph. The results of these
two types of initial conditions are highly correlated, with a Pearson
correlation coefficient of 0.999. The directional diffusion takes
longer, since the oil must spread along the diameter of the rock from
one end in this case, whereas in the anisotropic case there is already
oil spread (unevenly) throughout the rock at time $t=0$. On average,
the directional diffusion takes $1.44$ times longer than the anisotropic
one. Due to these similarities, hereafter only the case where the
whole rock is in contact with the reservoir is considered (Fig.~\ref{initial conditions}(b)).

It is considered that the diffusion process has taken place if $|u\left(i,t\right)-u\left(j,t\right)|\leq\delta$
for all pairs of nodes $i$ and $j$ in the graph when $t\rightarrow\infty$.
In this work a threshold of $\delta=10^{-5}$ is selected. This means
that if the ``concentration'' of the diffusive particle in one node
does not differ from that in any other node by more than $\delta=10^{-5}$
it is considered that the diffusion process has ended. Then, the time
at which this threshold is reached is recorded, and is called the
diffusion time. Due to the fact that many realizations of the same
process are carried out the average of this time is taken over all
these realizations. Once the values of the average time of diffusion
are obtained for each RFN, correlations are obtained with the structural
parameters considered in this work in order to see which of them influence
the diffusive process on the RFNs. Table \ref{correlation diffusion}
reports the Pearson correlation coefficients of these relations, where
the entries in boldface are those that are more significant from a
statistical point of view. The next subsection will analyze the theoretical
foundations for these findings.

\begin{table}[H]
\begin{centering}
\begin{tabular}{|c|c|c|c|c|c|c|}
\hline 
structural parameter  & correlation  & log  &  & structural parameter  & correlation  & log\tabularnewline
\hline 
\hline 
1  & 0.026  &  &  & $F_{1}$  & \textbf{ 0.815}  & Y\tabularnewline
\hline 
3  & -0.307  &  &  & $F_{2}$  & 0.415  & \tabularnewline
\hline 
4  & 0.225  &  &  & $F_{3}$  & \textbf{ 0.775}  & Y\tabularnewline
\hline 
5  & -0.613  &  &  & $F_{4}$  & \textbf{ 0.764}  & Y\tabularnewline
\hline 
6  & 0.142  &  &  & $F_{5}$  & 0.423  & \tabularnewline
\hline 
7  & -0.127  &  &  & $F_{6}$  & 0.418  & \tabularnewline
\hline 
8  & \textbf{0.809}  & Y  &  & $F_{7}$  & 0.387  & \tabularnewline
\hline 
9  & \textbf{0.745}  & Y  &  & $F_{8}$  & 0.419  & \tabularnewline
\hline 
10  & -0.495  & Y  &  & $F_{9}$  & 0.435  & \tabularnewline
\hline 
11  & 0.449  &  &  & $F_{10}$  & 0.416  & \tabularnewline
\hline 
12  & \textbf{-0.995}  & Y  &  & $F_{11}$  & 0.409  & \tabularnewline
\hline 
13  & \textbf{0.841}  & Y  &  & $F_{12}$  & 0.381  & \tabularnewline
\hline 
14  & 0.104  &  &  & $F_{13}$  & 0.424  & \tabularnewline
\hline 
15  & 0.328  &  &  & $F_{14}$  & 0.389  & \tabularnewline
\hline 
16  & \textbf{0.872}  &  &  & $F_{15}$  & 0.407  & \tabularnewline
\hline 
17  & \textbf{0.876}  &  &  & $F_{16}$  & 0.000  & \tabularnewline
\hline 
18  & \textbf{-0.845}  &  &  & $F_{17}$  & 0.412  & \tabularnewline
\hline 
19  & \textbf{0.923}  & Y  &  &  &  & \tabularnewline
\hline 
\end{tabular}
\par\end{centering}
\caption{Results of the regression analysis between the diffusion time in RFNs
and structural parameters of the same networks. The numerical values
correspond to the Pearson correlation coefficient and ``log'' indicates
whether the correlation is in a log-log scale. The numbers used for
the structural parameters are given in Table 1 and the structure of
the fragments is given in Appendix~\ref{appendix:small_subgraphs}.}

\label{correlation diffusion} 
\end{table}

Figure \ref{f;diff_diam} illustrates the correlation between the
average diffusion time obtained by simulation of the diffusive process
on the RFNs versus the same process simulated on the analogous RRNGs.
In addition, it provides evidence that the average diffusion time
simulated on the RFNs is correlated to some of the most important
structural parameters of the RRNGs. This means that the real-world
RFNs can be replaced by their analogous random models in order to
simulate the diffusion processes taking place on the rocks.

\begin{center}
\begin{figure}[h!]
\begin{centering}
\includegraphics[width=0.5\textwidth]{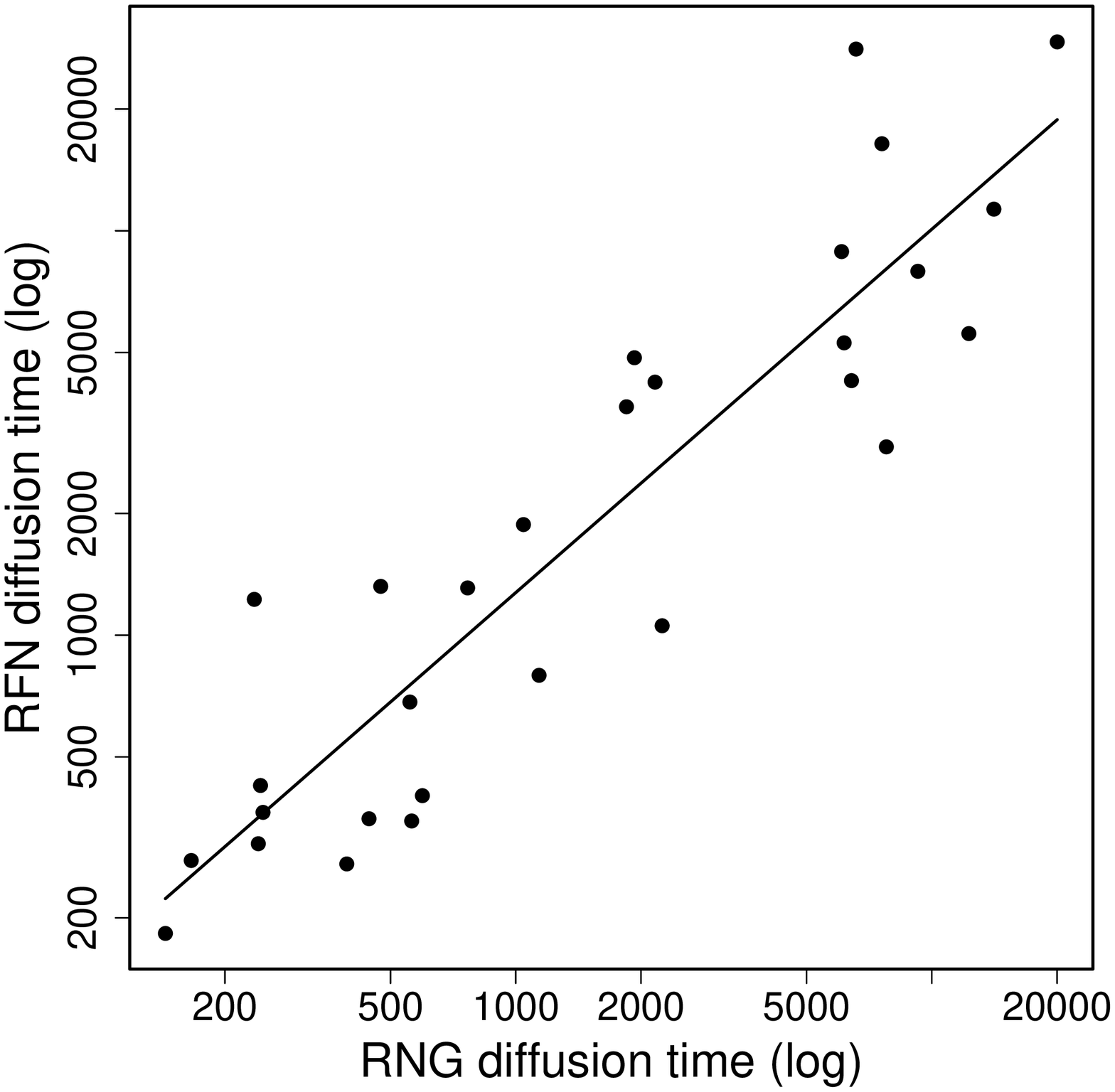}\includegraphics[width=0.5\textwidth]{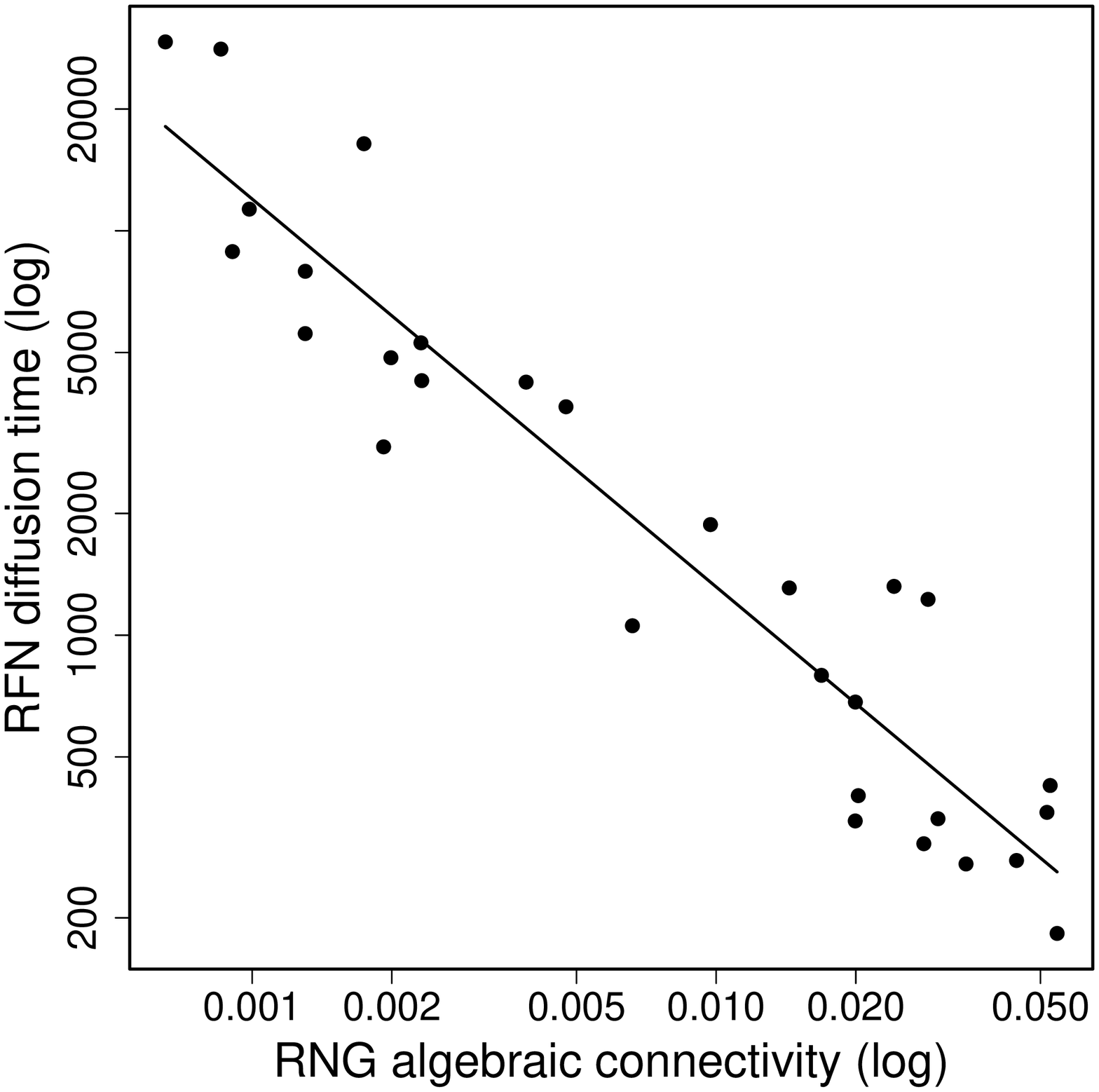}
\includegraphics[width=0.5\textwidth]{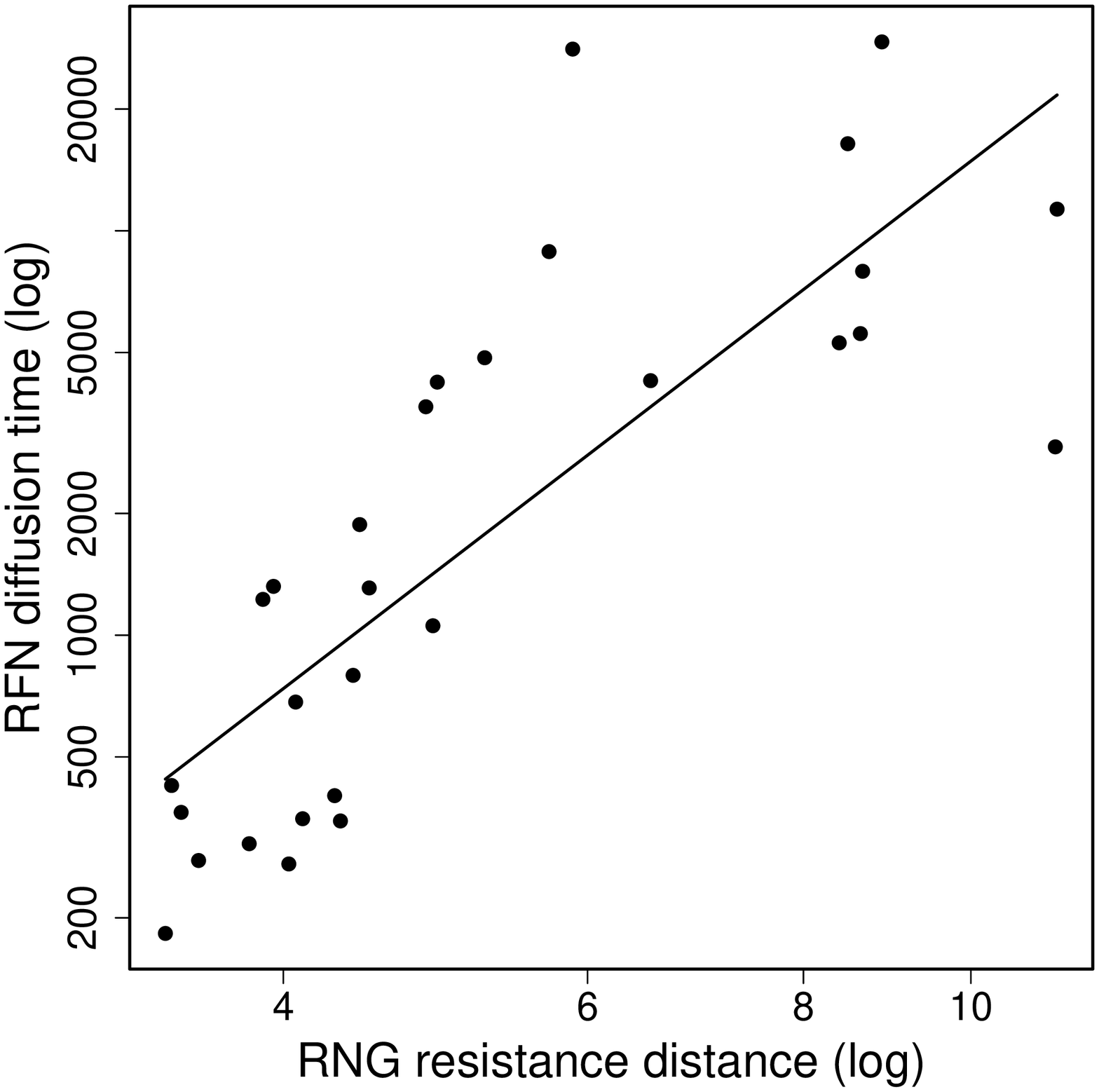}\includegraphics[width=0.5\textwidth]{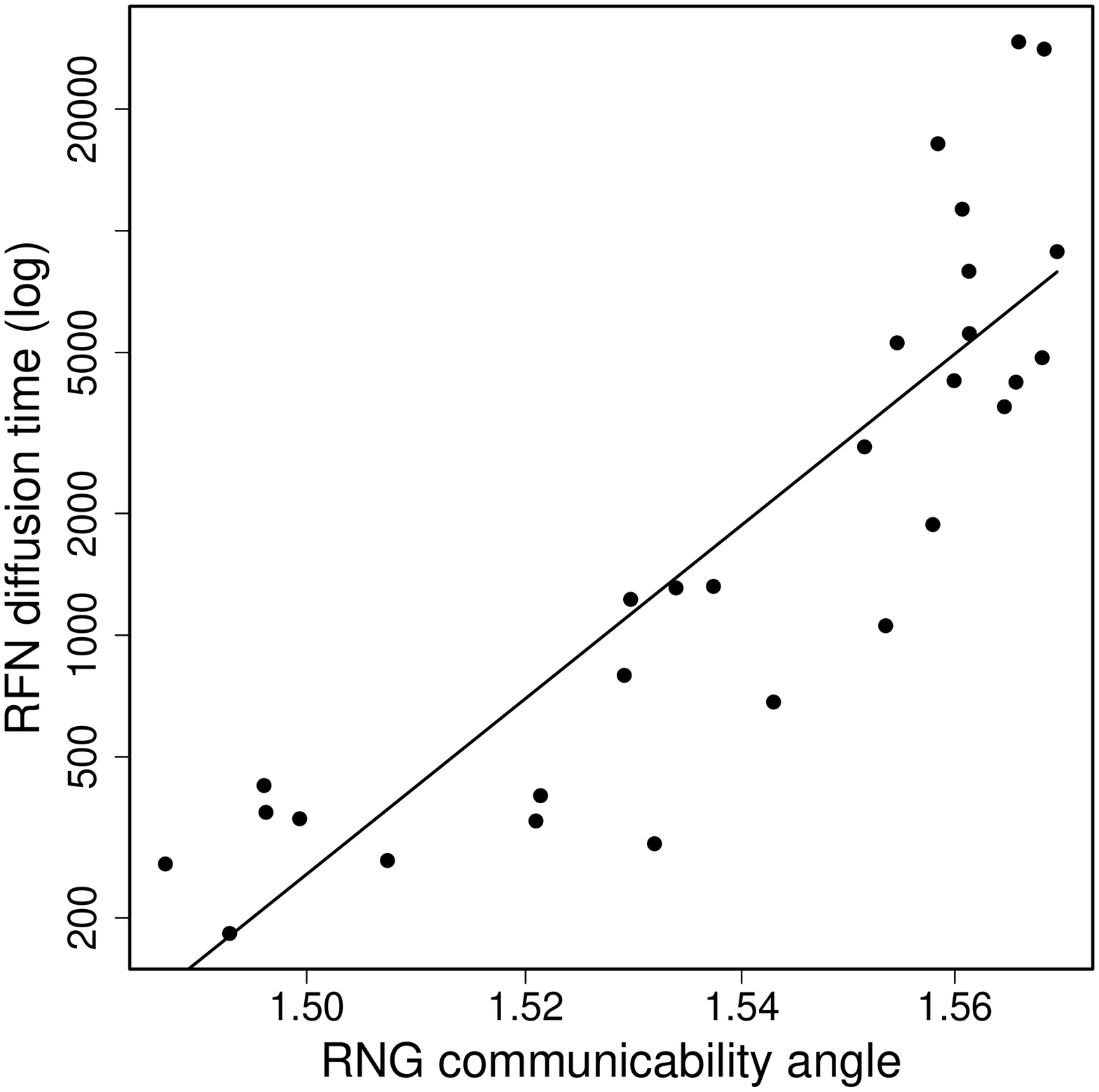} 
\par\end{centering}
\caption{(a) Plot of the average diffusion time of the RFNs versus the diffusion
time in the analogous RRNGs. Plot of the diffusion time of the RFNs
versus algebraic connectivity (b), resistance distance (Kirchhoff
index) (c) and average communicability angle (d) in the analogous
RRNGs.}
\label{f;diff_diam} 
\end{figure}
\par\end{center}

\subsection{Theoretical analysis}

As can be seen in Table (\ref{correlation diffusion}) the best structural
parameter describing the diffusion time is the second smallest eigenvalue
of the Laplacian matrix, $\mu_{2}$. In order to understand this relation,
consider the solution of the diffusion equation (Eq.~\ref{eq:diffusion equation}),
which is:

\begin{onehalfspace}
\begin{equation}
\vec{u}\left(t\right)=e^{-t\mathcal{L}}\vec{z},
\end{equation}

\end{onehalfspace}

where $\vec{z}=\vec{u}_{0}$. In terms of the spectral decomposition
of the graph Laplacian, the solution of the diffusion equation (Eq.~\ref{eq:diffusion equation})
can be written as

\begin{onehalfspace}
\begin{equation}
\vec{u\left(t\right)}=e^{-t\mu_{1}}\left(\vec{\psi}_{1}\cdot\vec{z}\right)\vec{\psi}_{1}+e^{-t\mu_{2}}\left(\vec{\psi}_{2}\cdot\vec{z}\right)\vec{\psi}_{2}+\cdots+e^{-t\mu_{n}}\left(\vec{\psi}_{n}\cdot\vec{z}\right)\vec{\psi}_{n},\label{eq:spectral consensus}
\end{equation}

\end{onehalfspace}

where $0=\mu_{1}<\mu_{2}\leq\cdots\leq\mu_{n}$ are the eigenvalues
and $\vec{\psi}_{j,p}$ the $p$th entry of the corresponding $j$th
eigenvector of the Laplacian matrix, and $\vec{x}\cdot\vec{y}$ represents
the inner product of the corresponding vectors. Equation \ref{eq:spectral consensus}
can be written for a given node $p$ as

\begin{onehalfspace}
\begin{equation}
\vec{u}_{p}\left(t\right)=\sum_{q=1}^{n}\vec{z}_{q}\sum_{j=1}^{n}\vec{\psi}_{j,p}\vec{\psi}_{j,q}e^{-t\mu_{j}},\label{eq:spectral one node}
\end{equation}

\end{onehalfspace}

which represents the evolution of the state of the corresponding node
as time evolves. Now, consider that the time tends to the time of
diffusion $t\rightarrow t_{c}$, where $t_{c}$ is the time at which
$\vert u_{i}(t)-u_{j}(t)\vert\leq\delta$ for all $i,j$. Denote this
time by $t_{c}^{-}$, and then

\begin{onehalfspace}
\begin{equation}
\vec{u}_{p}\left(t_{c}^{-}\right)=\frac{1}{n}\sum_{q=1}^{n}\vec{z}_{q}+\sum_{j=2}^{n}\left(\vec{\psi}_{j,p}e^{-t_{c}^{-}\left(p\right)\mu_{j}}\sum_{q=1}^{n}\vec{\psi}_{j,q}\vec{z}_{q}\right),\label{eq:near consensus}
\end{equation}

\end{onehalfspace}

where here $t_{c}^{-}\left(p\right)$ means the time at which the
node $p$ is close to reaching the diffusion state. Let $\left\langle \vec{u}_{0}\right\rangle =\frac{1}{n}\sum_{q=1}^{n}\vec{z}_{q}$
and write Eq.~\ref{eq:near consensus} as follows

\begin{equation}
\vec{u}_{p}\left(t_{c}^{-}\right)-\left\langle \vec{z}\right\rangle =\sum_{j=2}^{n}\left(\vec{\psi}_{j,p}e^{-t_{c}^{-}\left(p\right)\mu_{j}}\sum_{q=1}^{n}\vec{\psi}_{j,q}\vec{z}_{q}\right).
\end{equation}

A node $p$ is selected such that $\vec{\psi}_{2,p}$ has the same
sign as $\vec{\psi}_{2}\cdot\vec{z}$. Since $\mu_{2}$ is the smallest
eigenvalue in the sum on the right hand of the expression, this terms
tends to 0 slower than the terms for the other values of $j$. This
means that, if $\delta$ is sufficiently small, the values of $t_{c}$
and thus $t_{c}^{-}$ will be very large. Thus, it is possible to
ensure that the left side of the equation is small enough such that
$\sum\limits _{j=3}^{n}\left(\vec{\psi_{j,p}}e^{-t_{c}^{-}(p)\mu_{j}}(\vec{\psi_{j}}\cdot\vec{z})\right)<0$.
This implies that

\begin{onehalfspace}
\begin{equation}
\left(\vec{u}_{p}\left(t_{c}^{-}\right)-\left\langle \vec{z}\right\rangle \right)<\vec{\psi}_{2,p}e^{-t_{c}^{-}\left(p\right)\mu_{2}}\left(\vec{\psi}_{2}\cdot\vec{z}\right).
\end{equation}

Now, because $\left|\vec{u}_{p}\left(t_{c}^{-}\right)-\left\langle \vec{z}\right\rangle \right|\geq\delta$,

\begin{equation}
\delta\leq\left|\vec{u}_{p}\left(t_{c}^{-}\right)-\left\langle \vec{z}\right\rangle \right|<\left|\vec{\psi}_{2,p}e^{-t_{c}^{-}\left(p\right)\mu_{2}}\left(\vec{\psi}_{2}\cdot\vec{z}\right)\right|.
\end{equation}

Then, the time at which the diffusion is reached $t_{c}\left(p\right)$
is bounded by 
\begin{eqnarray}
t_{c}\left(p\right) & \geq t_{c}^{-}\left(p\right)\geq\frac{1}{\mu_{2}} & \ln\left|\frac{\vec{\psi}_{2,p}\left(\vec{\psi}_{2}\cdot\vec{z}\right)}{\delta}\right|.
\end{eqnarray}

\end{onehalfspace}

Finally, the average time of diffusion is bounded by

\begin{onehalfspace}
\begin{equation}
\left\langle t_{c}\right\rangle \geq\frac{1}{\mu_{2}n}\sum_{p=1}^{n}\ln\left|\frac{\vec{\psi}_{2,p}\left(\vec{\psi}_{2}\cdot\vec{z}\right)}{\delta}\right|.\label{eq:global time of consensus}
\end{equation}

\end{onehalfspace}

As can be seen from Eq.~\ref{eq:global time of consensus} the average
diffusion time in a network inversely correlates with the second smallest
eigenvalue of the Laplacian matrix (see also the negative Pearson
correlation coefficient in Table \ref{correlation diffusion}). This
analysis clearly indicates that $\mu_{2}$ can be used as an estimator
of the rate of diffusion of oil and gas in rock fracture networks.
Using our previous results in which we found a bound for the diameter
and for the algebraic connectivity of RRNGs we can easily prove the
following one.
\begin{lemma}
Let $G=G\left(n,a,\beta=2\right)$ be a connected RRNG with $n$ nodes
embedded in a rectangle of sides with lengths $a$ and $b=a^{-1}$.
Let $k_{max}$ be the maximum degree of any node in $G$. Then, the
second smallest eigenvalue of the Laplacian matrix of the RRNG is
bounded as 

\begin{equation}
\left\langle t_{c}\right\rangle \geq\left(\frac{a^{4}+1}{8k_{max}}\dfrac{1}{\log_{2}^{2}n}\right)\sum_{p=1}^{n}\ln\left|\frac{\vec{\psi}_{2,p}\left(\vec{\psi}_{2}\cdot\vec{z}\right)}{\delta}\right|.
\end{equation}
\end{lemma}
This result clearly explains why the diffusion time correlates relatively
well with the diameter of the graph. That is, as the elongation increases,
the algebraic connectivity decays as a consequence of the increase
of the diameter of the graph, which make the diffusion time increase.
In other words, increasing the elongation of the RRNGs makes the diffusion
process to take a longer time to finish.

A lower bound for the algebraic connectivity has also been reported
by \cite{Mohar} in terms of the average path length $\bar{l}\left(G\right)$
of the graph, which explains the correlation obtained between the
diffusion time and $\bar{l}\left(G\right)$

\begin{equation}
\mu_{2}\left(G\right)\geq\dfrac{4}{2\left(n-1\right)\bar{l}\left(G\right)-\left(n-2\right)}.\label{eq:bound path length}
\end{equation}

On the other hand, the high correlation observed between the diffusion
time and the so-called Kirchhoff index can also be understood by using
the relation in Eq.~\ref{eq:global time of consensus} because the
Kirchhoff index is defined by

\[
Kf=\sum_{i<j}\sum_{k=2}^{n}\dfrac{1}{\mu_{k}}\left(\psi_{k,i}-\psi_{k,j}\right)^{2},
\]
from which it can easily be seen that the largest contribution is
made by the second smallest eigenvalue of the Laplacian matrix $\mu_{2}$
and its corresponding eigenvector (the Fiedler vector $\vec{\psi}_{2}$).

The somehow unexpected relations are those observed between the diffusion
time and Estrada index, the entropy, the free energy and the average
communicability angle, which are based on the adjacency instead of
on the Laplacian matrix of the graph. These relations can be understood
through the structural interpretation of the Estrada index in term
of subgraphs of the graph. This index is defined as

\begin{eqnarray}
EE\left(G\right) & = & tr\left(A^{0}\right)+tr\left(A\right)+\dfrac{1}{2}tr\left(A^{2}\right)+\dfrac{1}{6}tr\left(A^{3}\right)+\dfrac{1}{24}tr\left(A^{4}\right)+\dfrac{1}{120}tr\left(A^{5}\right)+\cdots.
\end{eqnarray}

Clearly, $tr\left(A^{0}\right)=n$ and $\dfrac{1}{2}tr\left(A^{2}\right)=m$
(notice that the adjacency matrix is traceless due to the lack of
any self-loop in the networks). It is well-known that $\dfrac{1}{6}tr\left(A^{3}\right)=F_{2}$
is the number of triangles in the graph. In a similar way $tr\left(A^{4}\right)=2m+4F_{1}+8F_{5}$
and $tr\left(A^{5}\right)=30F_{2}+10F_{6}+10F_{8}.$ Consequently,
the Estrada index of a graph can be written as

\begin{eqnarray}
EE & = & n+\dfrac{13}{12}m+\dfrac{1}{6}F_{1}+\dfrac{5}{4}F_{2}+\dfrac{1}{3}F_{5}+\dfrac{1}{12}F_{6}+\dfrac{1}{12}F_{8}+\cdots,
\end{eqnarray}

which indicates that this index can be expressed as a weighted sum
of the number of small fragments in the graph. The correlation coefficients
between the diffusion time of the RFNs and their number of nodes and
edges are 0.863 and 0.846, respectively. Similarly, there are high
correlations with the fragments $F_{1}$, $F_{3}$, and $F_{4}$.
Because the RFNs are not very dense and as previously observed they
do not contain a large number of triangles (indeed there are a few
RFNs which are triangle-free), this can be approximated as

\begin{eqnarray}
EE & \approx & an+bm+cF_{1}+d,
\end{eqnarray}

where $a,b,c,d$ are coefficients. Indeed, a linear regression analysis
makes an estimation of these parameters as

\begin{eqnarray}
EE & \approx & 0.85n+0.93m+0.40F_{1}+1.56,
\end{eqnarray}
with a Pearson correlation coefficient $r>0.99999$ when including
all 29 RFNs studied here. Then, it can be concluded that the correlation
observed between the Estrada index and the diffusion time in RFNs
is due to the fact that the diffusion time is well described by a
few small fragments of the graphs, namely the number of nodes, edges
and paths of length 2 ($F_{1}$), which are the main contributors
to the Estrada index in these graphs. The correlations with $F_{3}$,
and $F_{4}$ observed in Table (\ref{correlation diffusion})) can
be easily explained by the fact that these fragments, in networks
of poor cliquishness, are related to the fragment $F_{1}$. Finally,
the relatively high correlations observed for the entropy, the free
energy and the average communicability angle can be explained the
fact that all of these measures are in some way related to the Estrada
index of the graph (see Table 1).

\section{Potential improvements of the model}

\subsection{Influence of fracture aperture}

Any model is always a simplification of the reality made on the basis
of a series of physical assumptions, empirical observations and availability
of mathematical and computational tools to solve it. In this work
we have used a few of these simplifications to produce a simple but
effective modeling of the diffusion of fluids through rock fracture
networks. However, there are a few areas in which improvements can
be implemented in order to gain more realistic description of the
physical processes taking place. One of our assumptions in this model
is that all the fractures have the same aperture. This assumption
is of course far from real, but it was used due to the lack of data
about such apertures for the rock fracture networks that we studied
in this work. However, such data is not difficult to obtain experimentally
and we will give here some hints about how to implement this parameter
and how it would affect the modeling results. 

When assuming that all fracture apertures were identical we used unweighted
graphs to represents the RFNs, i.e., every edge in the graph received
an identical unit weight. If information concerning the aperture of
the fractures were available we can represent it in our model by transforming
the graphs representing RFNs into weighted graphs, in which every
edge receives a weight corresponding to the aperture of that fracture.
As we do not have such data for the RFNs studied here we assume that
such apertures are randomly and independently distributed from $[0,1]$
over the fractures of a network. We then repeat our simulations for
these weighted graphs representing RFNs with different, randomly distributed,
apertures. We report here the average of the diffusion times for 10
random realizations.

In Figure~\ref{f;diff_weighted} we show the correlation between
the diffusion time in the weighted RFNs and the unweighted RFNs in
a log-log scale. It is clear that there is a strong correlation between
the two results, with a correlation coefficient of $0.870$. It is
obvious that the diffusion times will change in dependence of the
type of aperture that we use. For instance, the average times using
apertures in the range $[0,1]$ is larger than that when using all
weights equal to one. If apertures of larger magnitude were used,
an acceleration of the diffusion process should be observed, with
significantly smaller diffusion times that those observed for the
unweighted case. However, what is most important here is that a power-law
relation exists between the diffusion time in the unweighted and the
randomly weighted networks. That is, 

\begin{equation}
\left\langle t_{c}\right\rangle \sim\left\langle \tilde{t}_{c}\right\rangle ^{\kappa},
\end{equation}
where $\tilde{t}_{c}$ is the time of diffusion in the weighted RFN
and $\kappa\approx0.92$ is a fitting parameter obtained by using
nonlinear regression analysis for the data displayed in Figure~\ref{f;diff_weighted}.
More work is needed to show whether such kind of power-law relation
exists in general between these two parameters, which would indicate
some kind of universal scaling between the network with identical
apertures for all the fractures and that with apertures randomly and
independently distributed. In the meantime, we can assert that based
on the current results knowing the behavior of diffusion on the unweighted
RFNs also gives information about the behavior of the diffusion when
the aperture of the fractures is randomly and independently distributed.

\begin{center}
\begin{figure}[h!]
\begin{centering}
\includegraphics[width=0.65\textwidth]{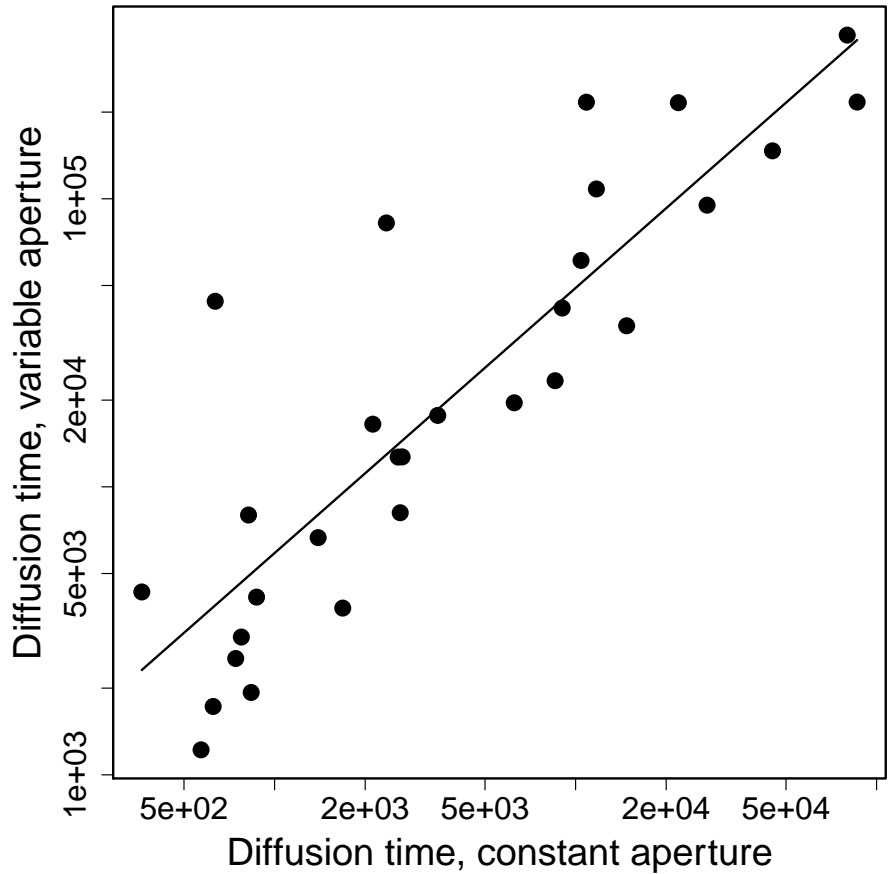}
\par\end{centering}
\caption{Plot of the diffusion time for the weighted RFNs against the diffusion
time with all edge weights equal to 1 (log-log scale).}
\label{f;diff_weighted} 
\end{figure}
\par\end{center}

\subsection{Influence of long-range hops of diffusive particles}

Another approximation that was considered in the current work is that
the conditions of the fracture networks in terms of their homogeneity
is appropriate for the use of the normal diffusion equation. However,
there is currently a vast volume of literature suggesting the lack
of homogeneity in many unconventional naturally fractured reservoirs
of oil \cite{1D_nanoporous_media}. These studies point out to the
existence of complex combinations of connected and isolated pores,
combinations of regions with discrete and continuous fractures and
variable properties of the hydrocarbon properties across the reservoir
\cite{velazquez_fractures}. These geological and petrophysical complexities
cannot be described by using the simple diffusion equation and much
more sophisticated models have emerged, which exploit the fractal
nature of such irregularities. It has been well documented that such
inhomogeneities in the properties of the systems to be modeled play
a major role in the diffusion processes taking place, which are quite
similar to anomalous diffusion in disordered media. Then, it is normal
to model the geostatistics of these reservoirs by a fractional Brownian
motion and fractional Lévy motion \cite{hewett_fractal,fractional_diffusion}.

Here we will follow a different path which connects in a natural way
with our previous model based on the normal diffusion equation on
graphs. We consider here a generalization of this equation using the
so-called $d$-path Laplacian operators introduced by \cite{estrada_multi_hop}.
In a recent work (\cite{estrada_superdiffusion}) have proved analytically
the existence of anomalous diffusion\textemdash superdiffusive and
ballistic behavior\textemdash when this model is used in infinite
one-dimensional systems. In the case of finite graphs, Estrada et
al. have shown that the biggest possible acceleration of diffusion
is obtained for certain parameters of the model in any graph, independently
of its topology.

Let us now define the $d$-path Laplacian matrices which account for
the hopping of the diffusive particle to non-nearest-neighbors in
the graph. Let $P_{l}\left(i,j\right)$ denote a shortest-path of
length $l$ between $i$ and $j$. The nodes $i$ and $j$ are called
the endpoints of the path $P_{l}\left(i,j\right)$. Because there
could be more than one shortest path of length $l$ between the nodes
$i$ and $j$ we introduce the following concept. The irreducible
set of shortest paths of length $l$ in the graph is the set $P_{l}=\left\{ P_{l}(i,j),P_{l}(i,r),...,P_{l}(s,t)\right\} $
in which the endpoints of every shortest-path in the set are different.
Every shortest-path in this set is called an irreducible shortest-path.
Let $d_{max}$ be the graph diameter, i.e., the maximum shortest path
distance in the graph. Now, we have generalized the Laplacian matrix
to the so-called $d$-path Laplacian matrices which are defined as
follows.
\begin{definition}
Let $d\leq d_{max}$. The $d$-path Laplacian matrix, denoted by $\mathcal{L}_{d}\in\mathbb{R}^{n\times n}$,
of a connected graph of $n$ nodes is defined as: 

\begin{equation}
\mathcal{L}_{d}\left(i,j\right)=\left\{ \begin{array}{r}
\delta_{k}\left(i\right)\\
-1\\
0
\end{array}\right.\begin{array}{l}
\textnormal{if \ensuremath{i=j} },\\
\textnormal{if node \ensuremath{d_{ij}=d} }\\
\textnormal{otherwise,}
\end{array}
\end{equation}
where $\delta_{k}\left(i\right)$ is the number of irreducible shortest-paths
of length $d$ that are originated at node $i$, i.e., its $d$-path
degree. 

We can now define the generalized diffusion equation in which the
$d$-path Laplacian operators are transformed by certain coefficients
that make that the hopping probability of the diffusive particle decay
with the distance that the particle is going to hop. Estrada et al.~have
analyzed mathematically three different transforms of the $d$-path
Laplacian operators and proved that for the infinite linear chain
there is superdiffusive behavior when the operators are transformed
by using the Mellin transform with $1<s<3$. Hereafter we adopt this
generalized diffusion equation which can be written in the following
way:
\end{definition}
\begin{eqnarray}
\dfrac{\partial}{\partial t}u\left(x,t\right) & = & -\left(\sum_{d=1}^{d_{max}}d^{-s}\mathcal{L}_{d}\right)u\left(x,t\right),\label{eq:generalised diffusion}\\
u\left(x,0\right) & = & u_{0}.\label{eq:initial condition_1}
\end{eqnarray}

Obviously, when $s\rightarrow\infty$ the terms $d^{-s}\rightarrow0$
for all $d>1,$ and we recover the normal diffusion equation (\ref{eq:diffusion equation}).
In this framework we should expect just the normal diffusion to take
place. However, when $s\rightarrow0$, the system behaves as a fully-connected
graph in which the diffusive particle can hop to any other node in
the RFN with identical probability. For the simulations, we consider
here only the case when $s=1$ and we use a stopping criterion of
$\varepsilon=10^{-4}$.

We have calculated the diffusion time using the generalized diffusion
equation (\ref{eq:generalised diffusion}) for all the RFN's studied
here and compare the generalized diffusion times with those of the
normal diffusion process. The normal diffusion time averaged for all
RFNs is $12,315$, while that for the generalized process is $0.69$.
If we consider the ratio of both times, i.e., the diffusion time under
the normal conditions $\left\langle t_{c}\right\rangle $ and the
diffusion time under long range hops of the diffusive particle $\left\langle \hat{t}_{c}\right\rangle $,
we see that on average it is $42,784$. It other words, the time for
diffusion on the RFNs decreases 40 thousand times when we consider
long-range hops. In Figure \ref{acceleration} we illustrate the ratio
$\left\langle t_{c}\right\rangle /\left\langle \hat{t}_{c}\right\rangle $
as the effect of long-range hops for the 29 real-world fracture networks
studied in this work. The most interesting thing in these results
is the fact that this ratio is dramatically larger than the average
for 3 of the RFNs studied. In these three cases the ratio $\left\langle t_{c}\right\rangle /\left\langle \hat{t}_{c}\right\rangle $
is ten times larger than the mean of this value for all the networks.
Although this is not a signature of superdiffusion it is worth further
investigation to determine whether superdiffusive behavior is observed
for these three networks under the long-range hop scheme. This, however,
is out of the scope of the current work and we leave it for a further
and more complete investigation of this phenomena. 

Another important characteristic of the results obtained in this subsection
of the work is the lack of correlation between $\left\langle t_{c}\right\rangle $
and $\left\langle \hat{t}_{c}\right\rangle $ (graphic not shown).
In contrast to what we have observed for the case of fracture aperture
where a power-law relation was observed between the time considering
random apertures and that using a fixed one, such a relation does
not exist here. This lack of correlation between $\left\langle t_{c}\right\rangle $
and $\left\langle \hat{t}_{c}\right\rangle $ could be indicative
of a different physical nature of the processes of diffusion under
normal conditions and the diffusion under long range hops of the diffusive
particle. This could show that the multi-hop approach to diffusion
captures some phenomenology not captured by the normal diffusion process,
which may include the case of superdiffusive behavior. 

\begin{center}
\begin{figure}[h!]
\begin{centering}
\includegraphics[width=0.55\textwidth]{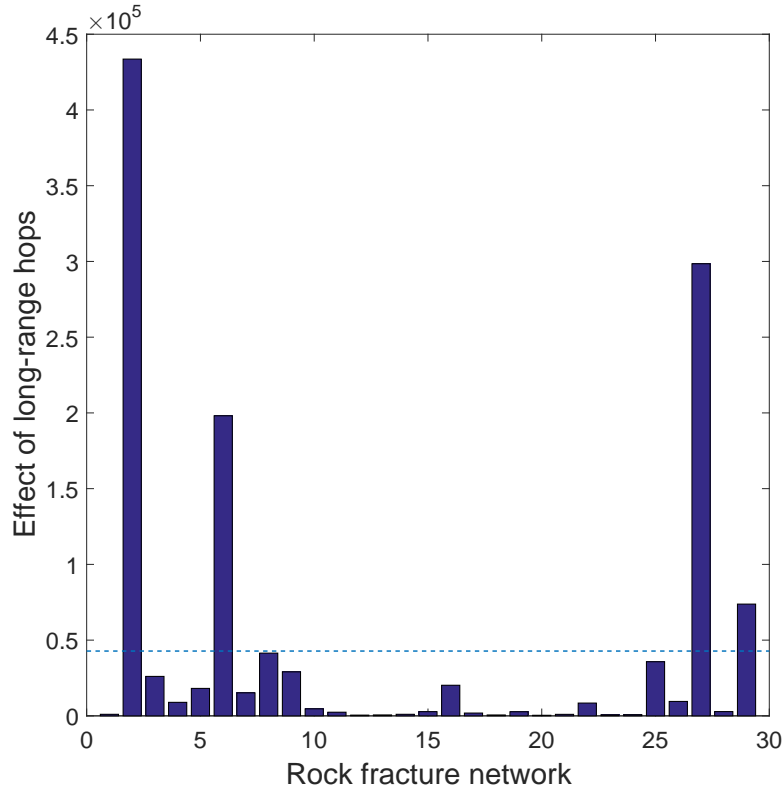}
\par\end{centering}
\caption{Bar plot of the values of the ratio $\left\langle t_{c}\right\rangle /\left\langle \hat{t}_{c}\right\rangle $
used to account for the effect of long-range hops on the 29 real-world
fracture networks studied in this work. The horizontal broken line
represents the average value of this ratio for the 29 RFNs. }
\label{acceleration} 
\end{figure}
\par\end{center}

\subsection{\textcolor{black}{Influence of 3D environments}}

Another characteristic of the current model that can be easily incorporated
to the study of diffusive processes is its extension to higher-dimensional
environments. We have considered here the 2-dimensional problem only
due to the fact that the data that we have represents the 2-dimensional
rock slices to investigate a projection of the fracture network into
a plane. However, it is obvious that such fracture networks cover
the 3-dimensional space of the rock and extend over its volume. 

\begin{figure}
\begin{centering}
\includegraphics[width=0.33\textwidth]{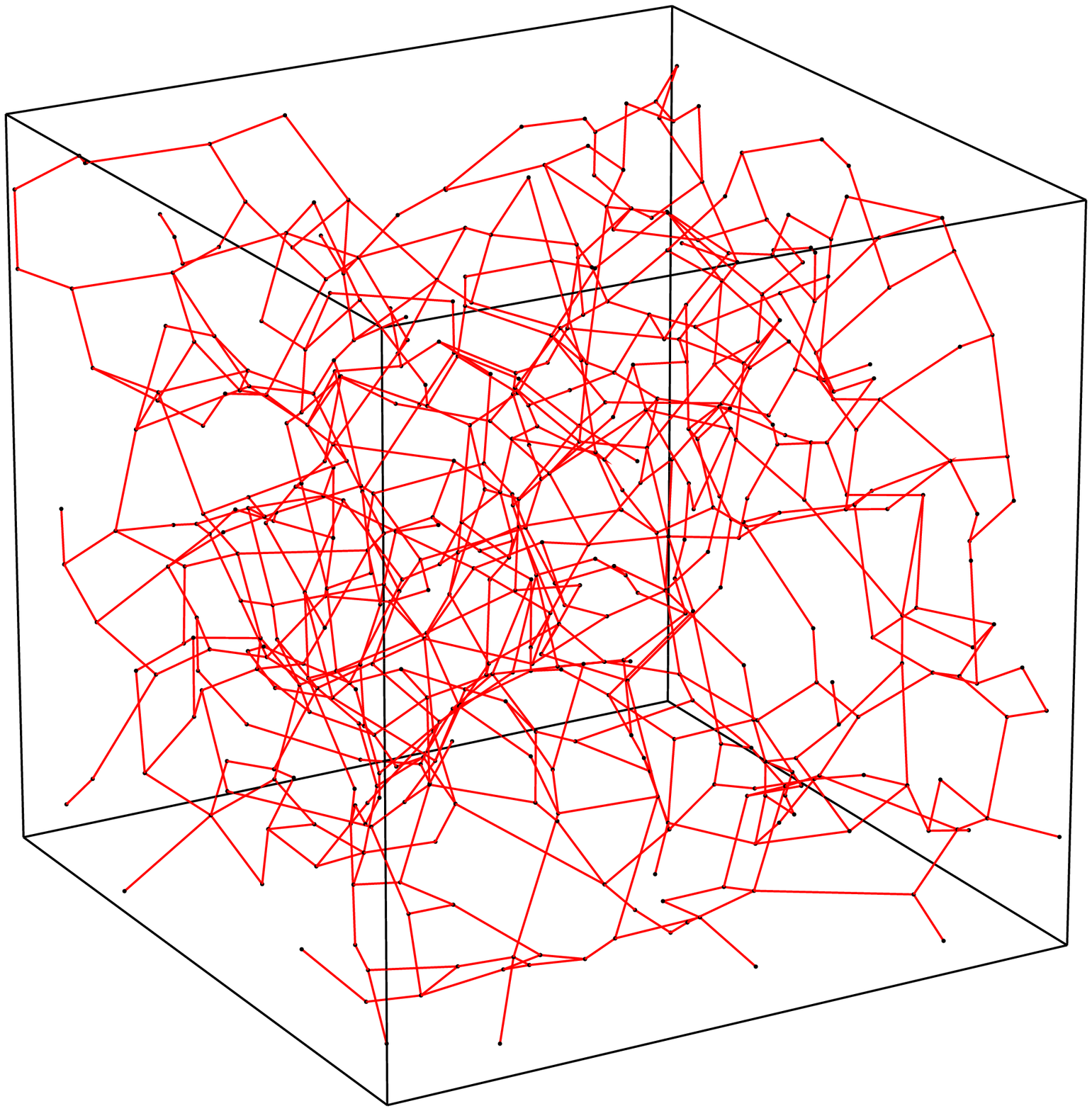}\includegraphics[width=0.33\textwidth]{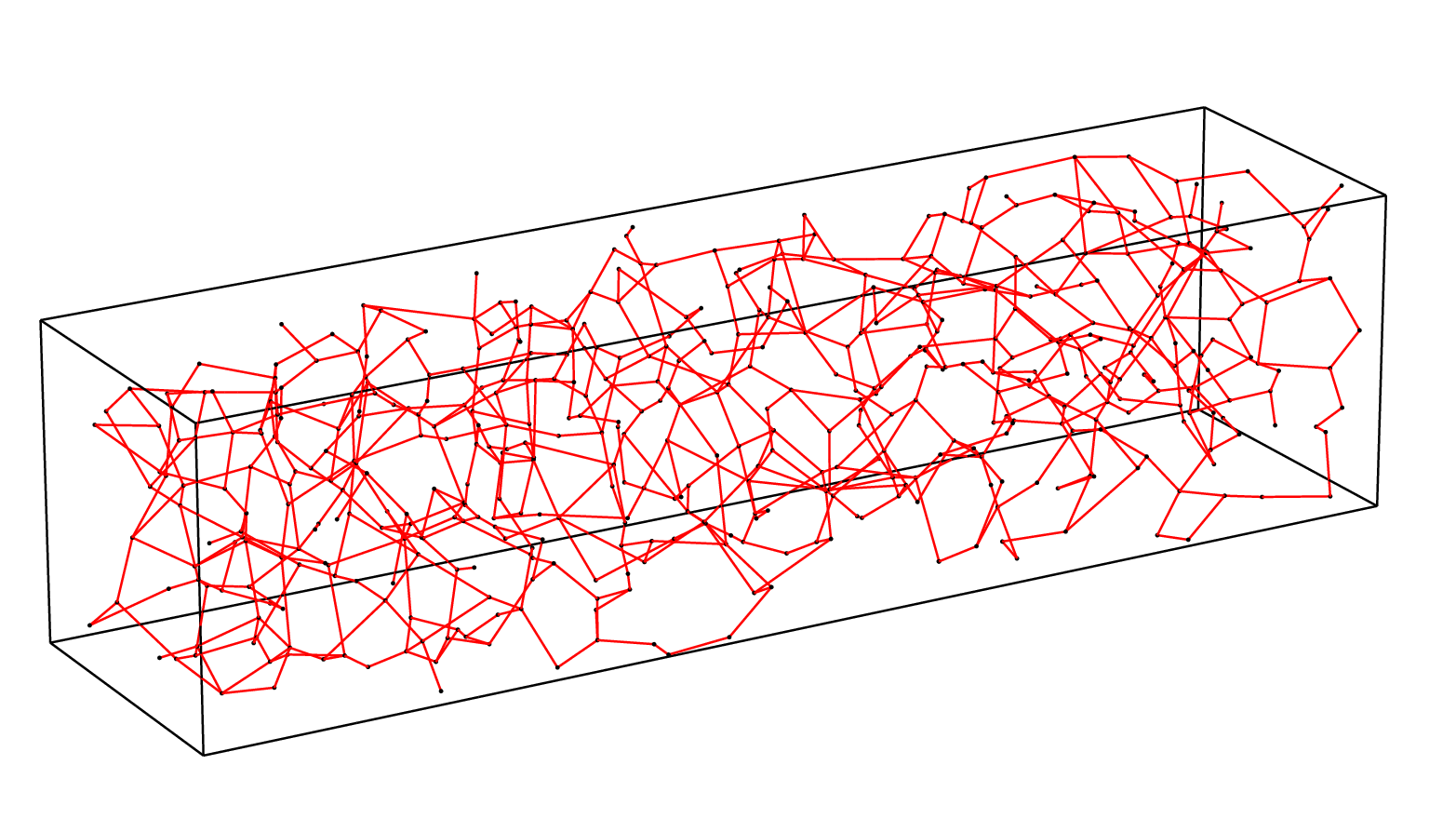}\includegraphics[width=0.33\textwidth]{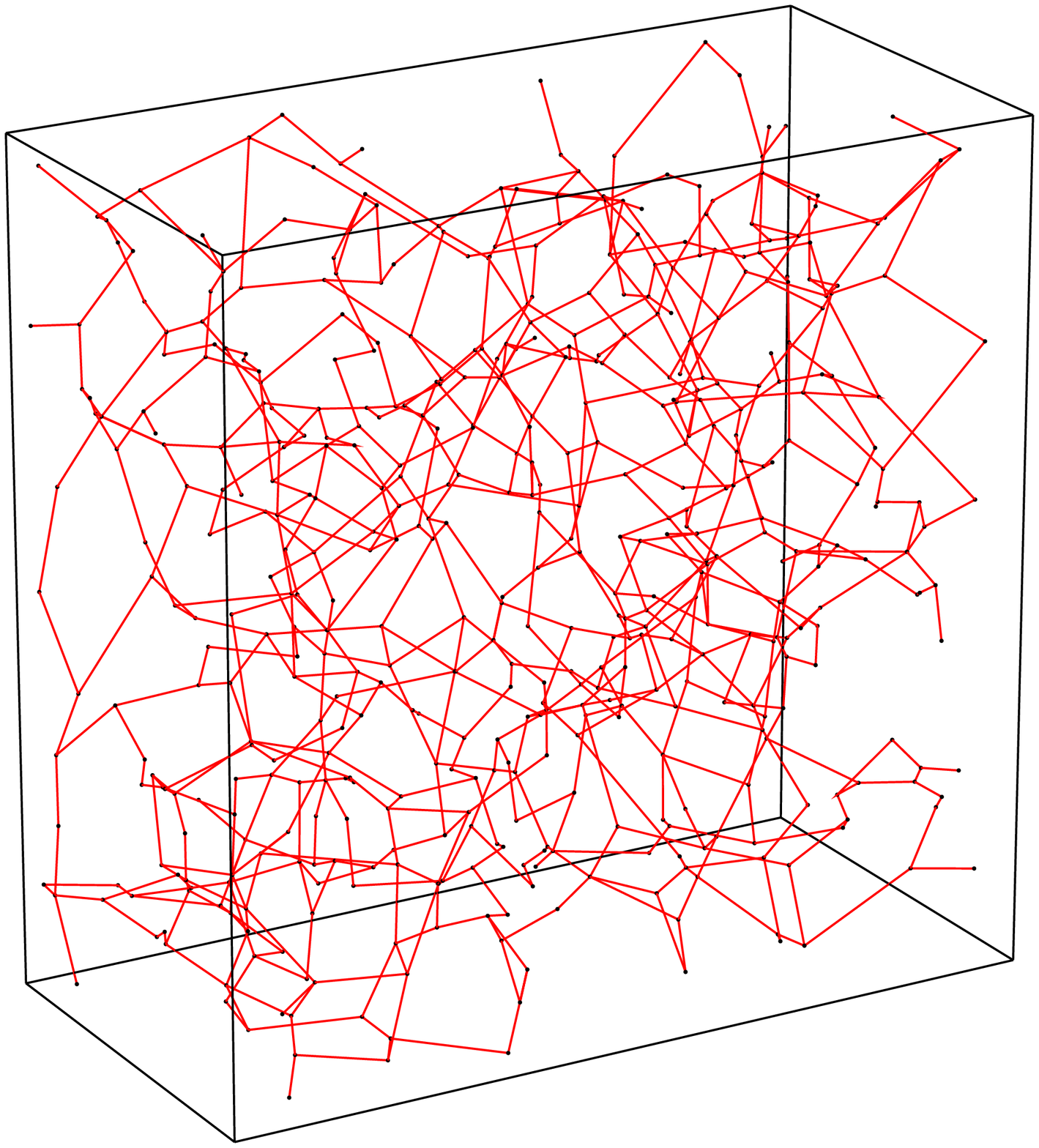}
\par\end{centering}
\caption{Illustration of some of the possibilities for modeling rock fracture
networks in 3-dimensional space.}

\label{3D cubes}
\end{figure}

For modeling purposes we should remark that now the number of parameters
increases and that more kinds of shapes emerge. While for the case
of the 2D scenario we can have only square-like or elongated rectangle-like
frameworks, in 3D we have the following three main possibilities of
environments: (i) a cube, (ii) an elongated cuboid, (iii) a flat cuboid.
We show them in Figure (\ref{3D cubes}). There are of course many
possible choices of the parameters in these models, which will cover
a large variety of shapes in 3D space. 

To give a flavor of the differences between the 2- and 3-dimensional
RRNGs we study here the change in the average degree, the diameter,
the algebraic connectivity and the diffusion time with the elongation
in the 2- and 3-dimensional RRNGs. To make things comparable we study
here only the case one side of the cube changes, say $1\leq a\leq5$
in steps for $0.5$, keeping b=1/a and $c=1$. As for the case of
the 2D graphs, the elongation of the cube produces a decay of the
average degree, an increase of the diameter and a decrease of the
algebraic connectivity. The resulting effects on the diffusion is
that elongation delays the diffusion process. It happens, as expected,
that the average degree of the 3D RRNG is larger than that of the
2D analogue. This is due to the fact that nodes can now be connected
in three directions of space instead of two. More interestingly, the
increase of the diameter of the 3D model is much slower than that
in the case of the 2D one. For instance, the diameter increases almost
linearly with the elongation according to $D\approx14.8+40.66a$ for
the 2D case, while it is $D\approx11.1+13.53a$ for the 3D case. That
is, the growth of the diameter in the 3D model studied is almost four
times slower than in the 2D case for similar elongations. These results
can be understood by adapting our previous bound for the 3D case.
In this case we should consider that the separation between the points
in the 3D cube is given by $1/\sqrt[3]{n}$, such as

\begin{equation}
D\left(G\right)\geq\sqrt[3]{n}\dfrac{\sqrt{\left(a^{4}+1\right)}}{a}.
\end{equation}

Then, it is clear that the diameter of the 3D RRNGs grows more slowly
than that of the 2D ones. This of course produces a dramatic decay
of the algebraic connectivity with the elongation in the 3D model
($\mu_{2}=0.0246a^{-1.87}$), where this parameter drops much faster
with the elongation than in the 2D case ($\mu_{2}=0.00307a^{-1.92}$).
The main consequence of this elongation effect is observed in the
slower growth of the diffusion time for the cuboid model than for
the rectangular one. While in the 2D case the diffusion time increases
as $\left\langle t_{c}\right\rangle \approx0.064+0.093a$ with the
elongation, in the 3D case it grows as $\left\langle t_{c}\right\rangle \approx0.059+0.028a$.
That is, the elongation of the cuboid seems to have an effect on the
diffusion time which is three times smaller than the effect of elongation
of the rectangle. 

\begin{figure}
\begin{centering}
\includegraphics[width=0.33\textwidth]{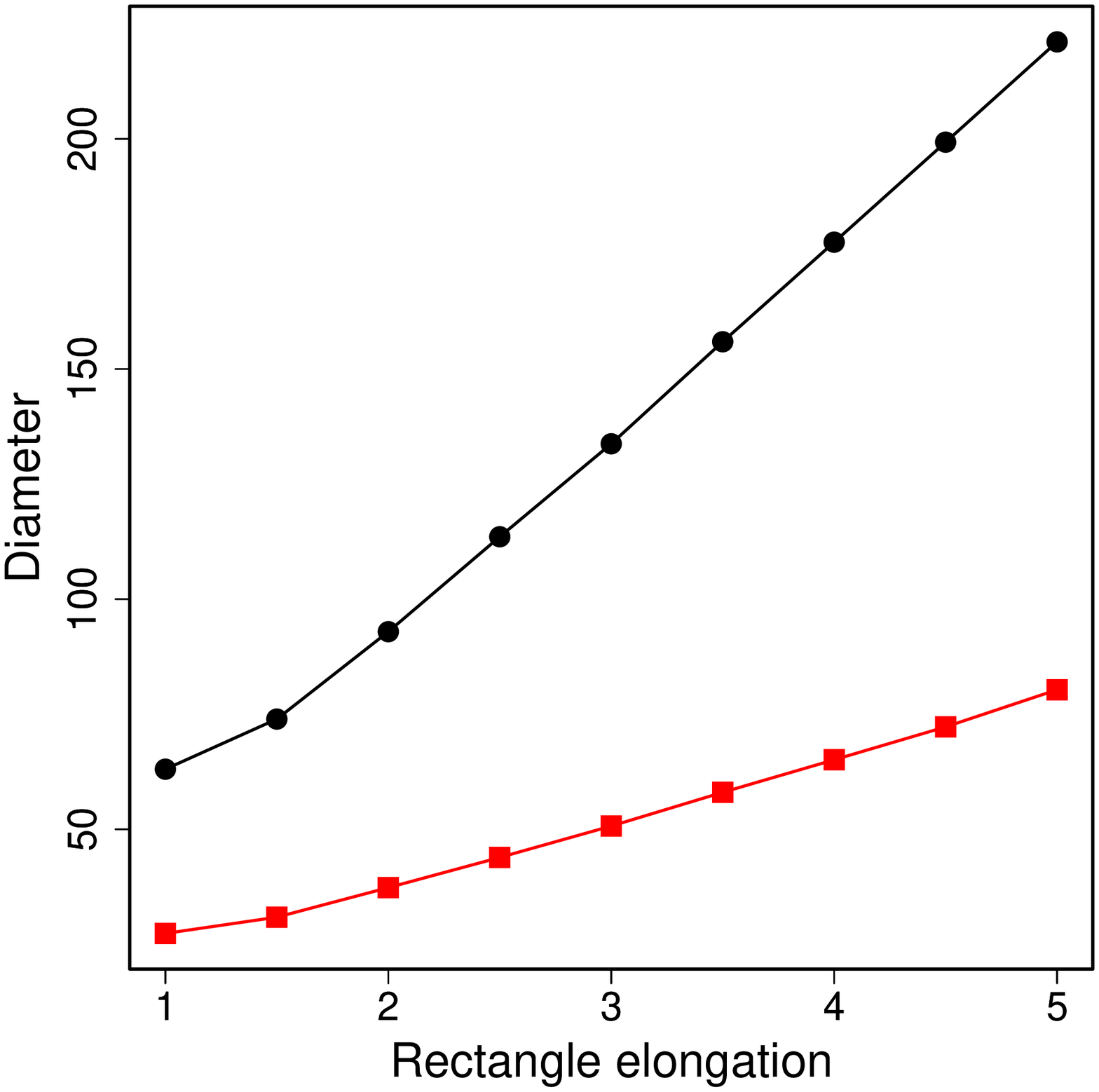}\includegraphics[width=0.33\textwidth]{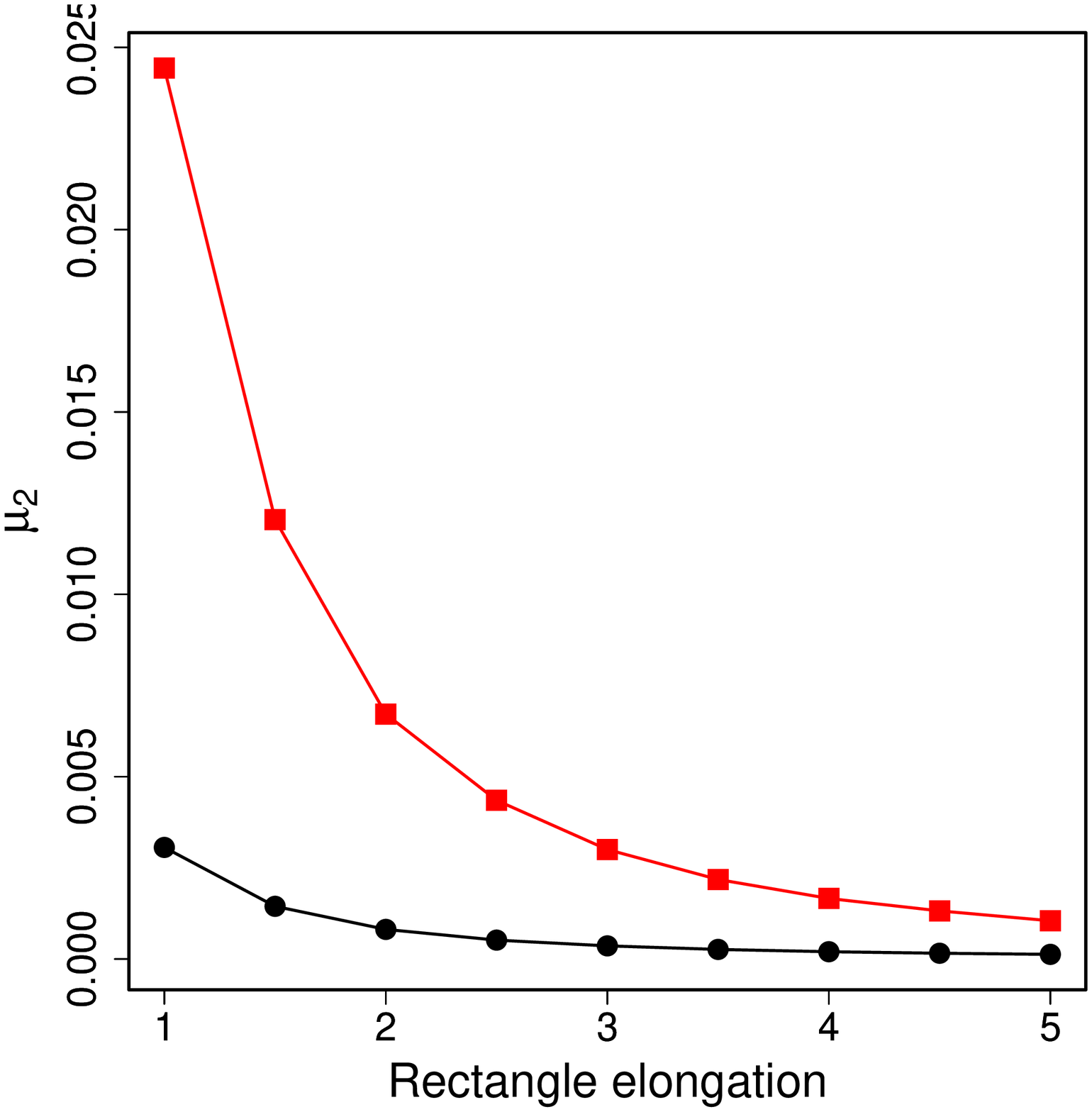}\includegraphics[width=0.33\textwidth]{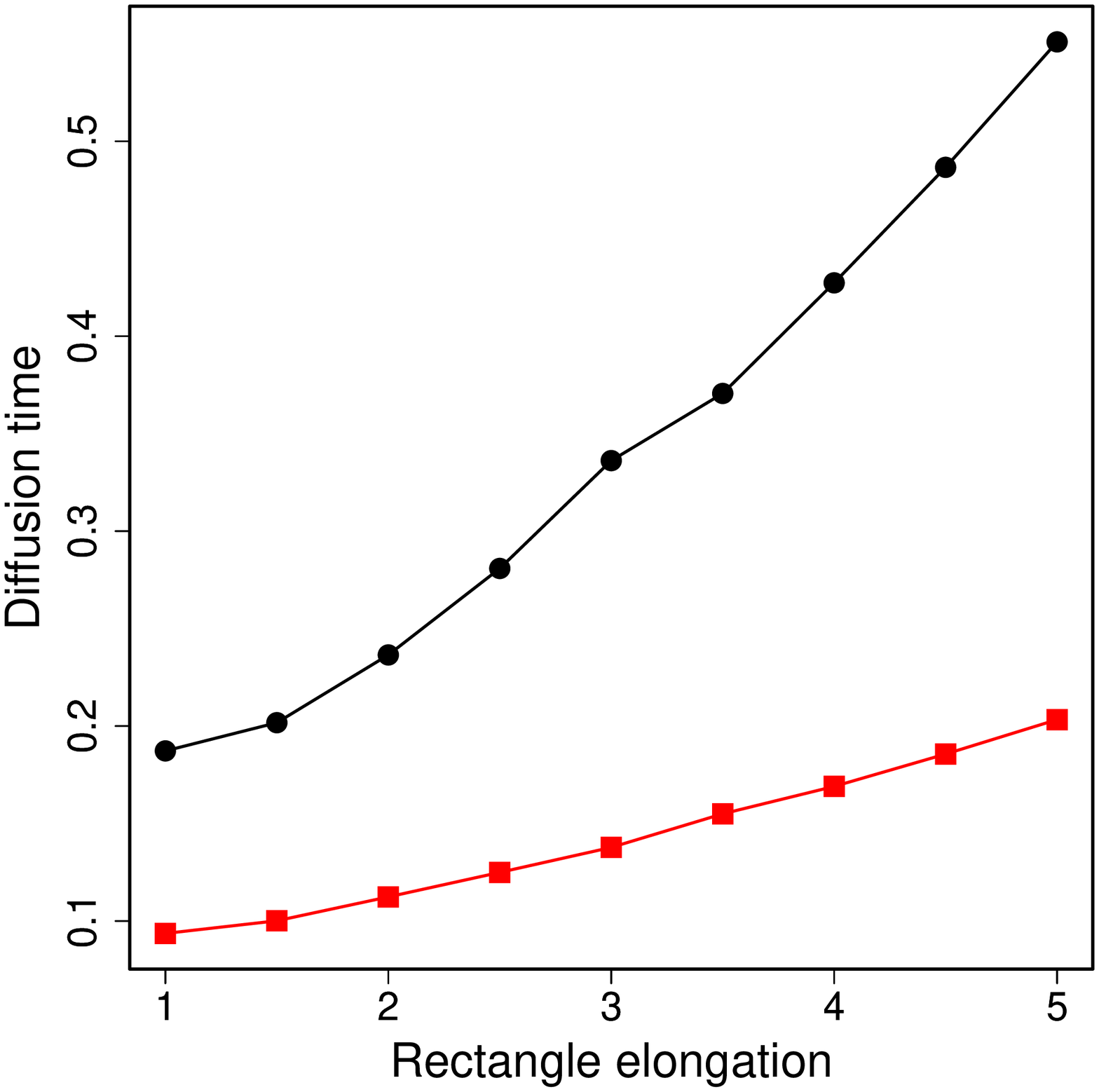}
\par\end{centering}
\caption{Comparison of the effects of elongation on the diameter, algebraic
connectivity and diffusion time in 2- (black dots) and 3-dimensional
(red squares) RRNGs having $n=1000$ nodes.}

\end{figure}

In closing we can think that the combination of the three effects
studied here as potential extensions of our model will make the diffusion
of oil and gas much faster than what we have predicted using the normal
2D diffusion model with constant apertures of the fractures. That
is, if we consider apertures larger than one, include long-range hops
of the diffusive particles and consider a 3D space instead of a 2D
one, we will observe super-fast diffusion on the rock fracture networks
studied. We should consider such combinations when some experimental
data is available, which permit us to compare our theoretical predictions
with reality.

\section{Conclusions}

We have developed a model based on a generalization of the random
neighborhood graphs which reproduces very well the structural and
dynamical properties of real-world rock fracture networks. The properties
of small rock fracture networks are well-described by using the newly
developed random rectangular neighborhood graph for small values of
the rectangular elongation. In contrast, larger RFNs are better described
by RRNGs with significantly longer elongations. This means that small
RFNs are embedded into more spherical rocks than the larger ones,
which are mainly embedded into rocks with a higher aspect ratio. The
most important characteristic of the RRNG is that it makes it possible
to study a large variety of structural and dynamical processes by
changing some of the parameters of the model. In such a way we can
be more independent of the existence of appropriate datasets of real-world
RFNs, which in many cases are scarce.

As has been seen in this work, the heat equation adapted to consider
the graph as the space in which the diffusive process is taking place
is an appropriate tool for studying the diffusion of oil and gas on
RFNs. This work also studies the relationship between the structure
of rock networks and the diffusion dynamics taking place on them.
In particular, a set of a few structural parameters were obtained
that describe well the diffusive process taking place on the networks.
Of special interest is the second smallest eigenvalue of the Laplacian
matrix, which shows a correlation coefficient larger than $0.99$
with the average diffusion time on RFNs. This index is then a very
good predictor of the capacity of a given network to diffuse substances
through the channels produced by the fractures in the rock.

Finally we have considered a few potential extensions of our model.
They include the consideration of variable fracture apertures, the
possibility of long-range hops of the diffusive particles as a way
to account for heterogeneities in the medium and possible superdiffusive
processes, and the extension of the model to consider 3-dimensional
modeling scenarios. These potential additions show the generality
and flexibility of our model to accommodate new structural and dynamical
characteristics to better reflect the physical reality to be modeled. 

All in all, we consider that the newly proposed model based on random
rectangular neighborhood graphs is very flexible and can be adapted
to specific simulation requirements and is therefore appropriate to
model RFNs in a variety of scenarios.

\section{Acknowledgment}

The authors thank E. Santiago et al. for providing the datasets of
rock fracture networks used in this work. Any request of this dataset
should be addressed to: esantiago@im.unam.mx. M.\ S.\ thanks financial
support from the Weir Advanced Research \foreignlanguage{american}{Centre}
at the University of Strathclyde. E.\ E.\ thanks the Royal Society
for a Wolfson Research Merit Award. EE also thanks Jorge X.Velasco-Hernández
for introducing him to this fascinating world of rock fracture networks.

\bibliographystyle{aps-nameyear}
\bibliography{Math_Geosci_paper_final}

\section*{Appendices}

\appendix

\section{ }

\label{appendix:small_subgraphs}

\begin{figure}[H]
\begin{centering}
\includegraphics[width=1\textwidth]{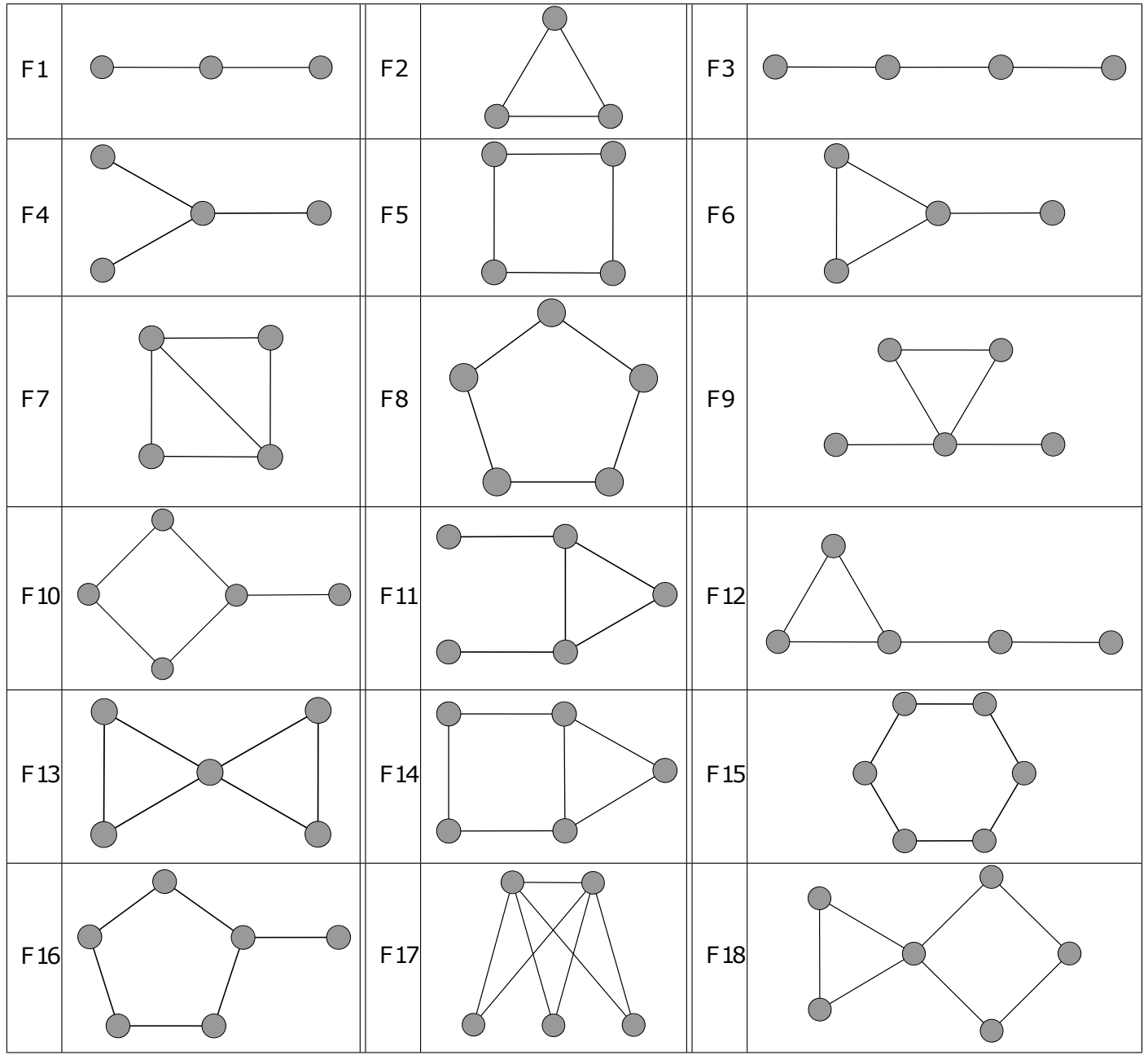} 
\par\end{centering}
\caption{Illustration of the structure of the small subgraphs used as structural
descriptors in this work.}

\label{small subgraphs} 
\end{figure}

\newpage{}


\section{ }

\label{appendix:small_subgraphs_code}

\begin{algorithm}
\begin{lstlisting}
A=A-diag(diag(A)); 
n=length(A); 
u=ones(n,1); 
t=diag(A^3); 
k=A*u; 
k1=k-1; 
k2=k-2; 
m=sum(k)/2;

%Auxiliary functions

Q=A^2.*A; 
P=0.5*A^2.*(A^2-1); 
q=diag(A^5); 
b=0.5*(q-5*t-2*(t.*k2)-2*Q*k2-2*(0.5*A*t-Q*u)); 
R=(1/6)*(A^2.*(A^2-1).*(A^2-2));

%Fragments according to book

F1=0.5*(k'*k1); 
F2=(1/6)*sum(t); 
F3=0.5*(k1'*A*k1)-3*F2; 
F4=(1/6)*(k.*k1)'*k2; 
F5=(1/8)*(trace(A^4)-4*F1-2*m); 
F6=0.5*(t'*k2); 
F7=0.25*u'*(Q.*(Q-A))*u; 
F8=(1/10)*(trace(A^5)-10*F6-30*F2); 
F9=0.25*(k2.*(k2-1))'*t; 
F10=u'*(P-diag(diag(P)))*k2-2*F7; 
F11=0.5*(k2'*Q*k2)-2*F7; 
F12=0.5*u'*(A^2-diag(diag(A^2)))*t-6*F2-2*F6-4*F7; 
F13=0.25*t'*(0.5*t-1)-2*F7; 
F14=0.5*u'*(Q.*(A^3.*A))*u-9*F2-2*F6-4*F7; 
F15=(1/12)*(trace(A^6)-2*m-12*F1-24*F2-6*F3-12*F4-48*F5-36*F7-12*F10-24*F13); 
F16=k2'*b-2*F14; 
F17=0.5*(u'*(R.*A)*u); 
F18=0.5*(u'*(P-diag(diag(P)))*t)-6*F7-2*F14-6*F17;

%Results in the form of a column vector

S=[F1 F2 F3 F4 F5 F5 F7 F8 F9 F10 F11 F12 F13 F14 F15 F16 F17 F18]'; 
\end{lstlisting}

\caption{Matlab code used for the calculation of the 18 small subgraphs illustrated
in Fig.\ (\ref{small subgraphs}).}
\end{algorithm}

\newpage{}

\section{ }

\label{appendix:beta_skeletons}

\begin{algorithm}
\begin{lstlisting}
function [graph,p] = beta(n,beta,lengths,p)
%Output 'graph' as adjacency list and list of coordinates 'p'
%can use cell2mat on output graph to get adjacency matrix
%
%beta=0 is the complete graph %beta=1 is the Gabriel graph
%beta=2 is the RNG
%
%If no lengths given, a unit square is assumed.
%If one length given, unit area is assumed.
%
%functions being called use beta differently, so the value is converted via
%beta_other=(1-beta)/(1+beta): [0,inf]->[1,-1]
%
%You can also supply the coordinates 'p' yourself as an n*2 matrix

if ~exist('beta','var')
    error('Please supply a value for beta')
end

if ~exist('n','var')
    error('Please supply the number of nodes')
end
if beta<0
    error('Beta must be non-negative')
end

if ~exist('lengths','var')
    lengths=[1 1];
elseif length(lengths)==1
    lengths=[lengths 1/lengths];
end

if ~exist('p','var')
    p=[unifrnd(0,lengths(1),n,1) unifrnd(0,lengths(2),n,1)];
end

if beta==0
    graph=CompleteGraph(n);
elseif beta==1
    dataDist=distFast(p,p);
    [~,indall]=sort(dataDist,'ascend');
    graph=GabrielGraph(p,indall);
elseif beta==2
    dataDist=distFast(p,p);
    graph=RelativeNeighborhoodGraph(dataDist);
else
    beta=(1-beta)/(1+beta);
    dataDist=distFast(p,p);
    graph=LuneBetaSkeletonGraph(p,dataDist,beta);
end
end 
\end{lstlisting}

\caption{Matlab code used for creating rectangular $\beta$-skeleton graphs,
which makes use of an available toolbox (\cite{proximity_toolbox}).}
\end{algorithm}

\end{document}